\documentclass[%
reprint,
amsmath,amssymb,
aps, nofootinbib,superscriptaddress
]{revtex4-2}
\usepackage{amsmath}
\usepackage{amssymb}
\usepackage{amsfonts}
\usepackage[utf8]{inputenc}
\usepackage[T1]{fontenc}
\usepackage{mathrsfs}
\usepackage{anyfontsize}
\usepackage[format=plain,justification=justified]{subcaption}
\usepackage{ragged2e}
\usepackage{url}
\usepackage[linktocpage,breaklinks]{hyperref}
\usepackage[usenames,dvipsnames]{xcolor}
\usepackage{txfonts}
\usepackage{tensor}
\usepackage{bm}
\usepackage{natbib}
\usepackage[linktocpage,breaklinks]{hyperref}
\usepackage{graphicx}
\usepackage{epsfig}
\usepackage{epstopdf}
\usepackage{natbib}
\usepackage{hyperref}

\hypersetup{colorlinks=true, citecolor=purple, linkcolor=purple,
	urlcolor=purple}
\usepackage{graphicx}
\usepackage{dcolumn}
\usepackage{bm}

\newcommand{\beq}{\begin{equation}}
\newcommand{\eeq}{\end{equation}}
\newcommand{\beqn}{\begin{eqnarray}}
\newcommand{\eeqn}{\end{eqnarray}}
\newcommand{\beqs}{\begin{subeqnarray}}
\newcommand{\eeqs}{\end{subeqnarray}}

\begin{document}
	\preprint{APS/123-QED}
	\title{Matter environments around black holes: Geodesics, light rings, and ultracompact configurations}
	
	\author{Dylan S. Fonseca}
	\email{dylan.fonseca@icen.ufpa.br}\affiliation{
		Programa de P\'{o}s-Gradua\c{c}\~{a}o em F\'{i}sica, Universidade Federal do Par\'{a}, 66075-110, Bel\'{e}m, Par\'{a}, Brazil
	}
    \author{Caio F. B. Macedo}
	\email{caiomacedo@ufpa.br}
	\affiliation{Faculdade de F\'{i}sica, Campus Salin\'{o}polis, Universidade Federal do Par\'{a}, 68721-000, Salin\'{o}polis, Par\'{a}, Brazil}
    \affiliation{
		Programa de P\'{o}s-Gradua\c{c}\~{a}o em F\'{i}sica, Universidade Federal do Par\'{a}, 66075-110, Bel\'{e}m, PA, Brazil
	}
    \author{Mateus Malato Corrêa}
    \email{malato.mateus@gmail.com}
    \affiliation{
		Programa de P\'{o}s-Gradua\c{c}\~{a}o em F\'{i}sica, Universidade Federal do Par\'{a}, 66075-110, Bel\'{e}m, PA, Brazil
	}
    \author{Diego Rubiera-Garcia}
	\email{drubiera@ucm.es}
	\affiliation{Departamento de Física Teórica and IPARCOS,
Universidad Complutense de Madrid, E-28040 Madrid, Spain
	}
	\date{\today}

    \begin{abstract}
Astrophysical black holes are invariably embedded in matter environments whose gravitational influence can alter key strong-field features of the spacetime. In this work, we investigate the impact of spherically symmetric dark-matter distributions on black hole geometry, geodesic structure, and ringdown phenomenology. Modeling the surrounding matter through Einstein clusters, we construct self-consistent spacetimes for three widely used density profiles—the Hernquist, Navarro–Frenk–White (NFW), and Jaffe models—and examine how their near-horizon behavior modifies the location and stability of circular timelike and null geodesics, including the innermost stable circular orbit (ISCO) and light rings. In the low-compactness regime, we derive analytical expressions showing that environmental effects generically shift the ISCO inward and the principal light ring outward, leading to parametric deviations in their associated orbital frequencies and Lyapunov exponents. At higher compactness, we explore the emergence of additional light rings, marginally stable orbits, and secondary horizons, identifying the regions of parameter space in which these additional geodesic structures arise. We introduce a perturbative approach to find approximate analytical solutions to the spacetime metric that are valid in astrophysical scenarios, and compare them with the numerical solutions. Using time-domain evolutions of scalar perturbations, we demonstrate how such structures can imprint characteristic signatures on the ringdown signal, including long-lived trapped modes and echo-like modulations associated with multiple potential barriers. Our results provide a unified framework for assessing environmental effects around black holes and highlight the importance of matter-induced corrections for interpreting upcoming electromagnetic and gravitational-wave observations.
\end{abstract}
\maketitle

\section{Introduction}\label{sec:Introduction}

Black holes (BHs) stand as one of the most paradigmatic solution in General Relativity (GR), embodying the theory’s remarkable simplicity. Being fully described  in terms of only three macroscopic quantities—mass, charge, and angular momentum—they represent the final outcome of full gravitational collapse and play a central role in our understanding of strong-field gravity. In vacuum, the Kerr family of solutions \cite{Newman:1965tw,Teukolsky:2014vca} exhausts the possibilities for stationary, asymptotically flat BHs, a simplicity often referred to as the ``no-hair'' property \cite{Chrusciel:2012jk}. However, astrophysical BHs are not isolated: they reside in galaxies filled with stars, plasma, magnetic fields, and dark matter (DM). Environmental effects around these objects therefore demand extending the vacuum paradigm to include surrounding matter distributions, leading to spacetimes that deviate—sometimes subtly, sometimes significantly—from the pure Kerr geometry \cite{Cardoso2022,Figueiredo2023,Maeda2024} and which may hamper our ability to test strong-field gravity \cite{Barausse:2014tra,Cole:2022yzw}.

DM is believed to pervade the Universe, constituting most of its matter content, and is expected to accumulate around compact objects through dynamical relaxation or adiabatic accretion processes \cite{Bertone:2002je,Guver:2012ba,Bell:2019pyc,Davies:2019wgi,Kavanagh:2020cfn}. In the galactic context, supermassive BHs naturally reside at the centers of DM halos, potentially giving rise to dense cusps or spikes near the horizon \cite{Ullio:2001fb,Alvarez:2020fyo,Liu:2024xcd}. The structure and density of these halos are typically described by empirical or simulation-driven profiles. Among the most widely used are the Hernquist \cite{Hernquist1990}, Navarro–Frenk–White (NFW) \cite{Navarro1996}, and Jaffe \cite{Jaffe1983} models, which differ mainly in their inner slope and asymptotic behavior. The Hernquist profile reproduces the de Vaucouleurs law for galactic bulges, the NFW distribution arises from cosmological N-body simulations, and the Jaffe model captures steep inner cores. Close to a BH, these profiles are expected to be modified: relativistic effects and accretion lead to vanishing densities at the horizon, but possibly large gradients in the near-horizon region \cite{Peirani:2008bu,Ferrer:2017xwm,GondoloSilk1999,Sadeghian2013,Xiong2025}.

A detailed understanding of such configurations is essential for interpreting electromagnetic and gravitational-wave (GW) signals from compact objects \cite{Ellis:2017jgp,Bezares:2019jcb,Speeney2022,Boudon:2023vzl,Haroon2025,Zhang2021,Zhang2022}. In particular, the geometry surrounding the BH affects the structure of geodesics \cite{Sharipov2025,Igata2023,Harada2023}, the location of the innermost stable circular orbit (ISCO), and the position of light rings (LRs)---unstable circular null geodesics that play a key role in high-frequency GW emission, lensing, and shadow formation \cite{Cardoso2022, Figueiredo2023}---. These same null unstable orbits underpin the interpretation of the quasinormal mode (QNM) spectrum of perturbed BHs, as the real and imaginary parts of the dominant ringdown frequencies are closely related— in the eikonal limit— to the LR frequency and the Lyapunov exponent, respectively \cite{Cardoso2014,Glampedakis:2019dqh,Pedrotti:2025idg}. Extensions involving surrounding matter modify these relations \cite{Chen:2023akf}, in particular introducing small redshifts or even creating new, long-lived oscillations when additional trapping regions emerge. Indeed, ultracompact objects (UCOs) or composite systems with stable LRs can support quasi-stationary modes with extremely long decay times, potentially giving rise to observable late-time signals distinct from standard BH ringdown \cite{Cardoso2016, Cardoso2022}.

Furthermore, in dense DM environments, the coexistence of compact matter configurations and central BHs opens the possibility of \emph{parasitic BHs}: small horizons nested within larger, self-gravitating dark structures \cite{Maeda2024}. Such systems blur the line between BHs and alternative compact objects, suggesting that DM halos themselves might become ultracompact and harbor central horizons. Understanding their spacetime structure and stability is crucial for connecting theoretical predictions to the phenomenology of GWs and BH shadows \cite{Cardoso:2022nzc, Bamber:2022pbs,Zhong:2023xll,Aurrekoetxea:2023jwk,Aurrekoetxea:2024cqd,Figueiredo:2023gas}.

The main objective of this work is to investigate the dynamics of both timelike and null geodesics in BHs surrounded by DM. Our analysis focuses on the three representative density profiles mentioned above: the Hernquist, Navarro–Frenk–White (NFW), and Jaffe models. We examine how these matter distributions modify the geodesic structure of the spacetime, a key aspect for understanding emission processes from accretion disks and GW sources. In particular, we discuss the modifications to the location and stability of both circular and timelike geodesics, most notably for the physically relevant cases of the innermost stable circular timelike geodesic orbit (ISCO) and the LRs. We provide ready-to-use expressions in the low-compactness regime, relevant for BHs embedded in low-density DM environments. In addition, we explore the high-density regime, discussing possible implications for the formation of ultracompact DM stars hosting parasitic BHs. 

This work is organized as follows. In Sec. \ref{S:II} we introduce the anisotropic matter configurations we are interested in, cast the field equations, and particularize them to the case of Einstein clusters, defined by vanishing radial pressure, and review the features of the three density profiles discussed above. The geodesic structure of the class of space-times with arbitrary mass functions is discussed in Sec. \ref{S:IV}, focusing on the existence, characterization, and stability of circular geodesics, particularizing these results for the case of low-compactness configurations. Original expressions for the locations of the LR and the ISCO in the Jaffe and NFW models within the low-compactness regime are derived. High-compactness configurations are fully considered in Sec. \ref{S:V}, constraining the space of parameter for the three density profile models in which new horizons, LRs, and marginally-stable orbits emerge. Analytical and numerical solutions are discussed in Sec. \ref{S:VI}, where we have employed a simple perturbative approach in order to find approximate analytical solutions to the spacetime metric, which we named Post-Schwarzschild expansion. We discuss the associated GW ringdown phenomenology including additional modes and echoes in Sec. \ref{S:VII}. Finally, Sec. \ref{S:VIII} contains the conclusion of our work and final thoughts. Throughout this work, we adopt the metric signature $(-,+,+,+)$ and natural units with $G=c=1$.

\section{Anisotropic matter configurations, Einstein clusters and density profiles} \label{S:II}

In order to find the spacetime geometries produced by BHs surrounded by density profiles configurations we need to account for the self-gravity of the mass distributions. It is known that scalar and electromagnetic fields can be effectively described by an anisotropic fluid as sources of gravity \cite{Boonserm2016}, and such sources are furthermore flexible enough to account for a variety of metrics proposed in the literature \cite{Bronnikov:2021uta}. This fact motivates us to consider the following energy-momentum tensor
\begin{equation}
    {T^\mu}_\nu=\text{diag}(-\rho,p,q,q),
    \label{em}
\end{equation}
where $\rho(r)$ is the energy density, $p(r)$ is the radial pressure and $q(r)$ is the tangential pressure. Throughout this work we shall make the assumption that the matter configuration exhibits spherical symmetry and, as a consequence, all of the energy-momentum tensor components are functions of $r$ only.

We also need to assume a static and spherically symmetric $ansatz$ for the metric, given by
\begin{equation}
    ds^2=-f(r)dt^2+\frac{dr^2}{h(r)}+r^2d\Omega^2,
    \label{metric}
\end{equation}
where $d\Omega^2=d\theta^2 + \sin^2 \theta d\varphi^2$ is the metric on the unit 2-sphere, the functions $f(r)$ and $h(r)=1-2m(r)/r$ are to be determined by the Einstein field equations ${G^\mu}_\nu=8\pi {T^\mu}_\nu$, and $m(r)$ is conveniently introduced as the mass function. Since we are only interested in asymptotically flat spacetimes, both metric functions $f(r)$ and $h(r)$ must suitably converge to unity as the radial coordinate gets large enough.

The $tt$ and $rr$ components of the field equations can be put in the form of the following coupled system
\begin{align}
    m'&=4\pi r^2\rho\label{m},\\
    \frac{f'}{f}&=\frac{2(m+4\pi r^3 p)}{r(r-2m)}\label{f}.
\end{align}
The above field equations need to be supplemented by the energy-momentum tensor conservation equation $\nabla_{\mu}T^{\mu\nu}=0$, which reads as
\begin{equation}
    p'=-\frac{f'}{2f}(\rho+p)-\frac{2}{r}(p-q)\label{p}.
\end{equation}
Observe that the contribution of the anisotropy assumption is contained precisely in the last term of the above equation. In contrast with the isotropic case, where only one equation of state is needed in order to completely determine the system, the anisotropic model generically needs to be supplemented by a pair of equations of state in order to form a complete system of Tolman-Oppenheimer-Volkoff (TOV) equations that determine the spacetime and the matter variables \cite{Bowers1974}.

A simplified approach for modeling these matter distribution is to assume that the matter particles are moving in circular geodesics under the influence of their collective gravitational field, without any interaction between each another. From the composition of many of such shells, each with a well defined radius and angular momentum, emerges an anisotropic fluid without any radial pressure (since the centrifugal force is exactly balanced by the gravity pull), but with an energy density $\rho(r)$ and a tangential pressure $q(r)$. The result of this construction is called an Einstein cluster \cite{Einstein1939,GeralicoPompiRuffini2012}, and has been previously shown to be capable of reproducing galactic rotation curves \cite{BohmerHarko2007,Jusufi:2022jxu,Maeda:2024tsg}. Its energy-momentum tensor is thus given by
\begin{equation}
    {T^\mu}_\nu=\text{diag}(-\rho,0,q,q).
    \label{ec}
\end{equation}
Therefore, by imposing $p=0$ in Eqs.~\eqref{m}--\eqref{p}, we arrive at the following system of field equations
\begin{align}
    m'&=4\pi r^2\rho\label{eqm},\\
    \frac{f'}{f}&=\frac{2m}{r(r-2m)}\label{eqf},\\
    2q&=\frac{\rho m}{r-2m}\label{eqq}.
\end{align}

We have therefore constructed a setup in which once a density profile function $\rho(r)$ is provided, the system of equations above is closed and one should be able to determine the functions $f(r)$, $m(r)$ and $q(r)$, given adequate boundary conditions. Our next goal is thus to consider specific choices for such a density profile.

\subsection{Density profiles around black holes} \label{S:III}

We consider a matter distribution described by the following density profile 
\begin{equation}
    \rho(r)=\frac{\rho_0(1-r_\text{in}/r)^\delta}{(r/a_0)^\gamma(1+(r/a_0)^\alpha)^{(\beta-\gamma)/\alpha}}.
    \label{rho}
\end{equation}
Observe that this generic model is determined by four parameters $(\alpha,\beta,\gamma,\delta)$, where $\beta$ and $\gamma$ characterize the asymptotic power-law behaviors of the profile in the so-called far zone ($r\gg a_0$) and in the near zone ($r\ll a_0$) respectively, with a typically large length scale $a_0$, whereas the parameter $\alpha$ describes the transition-region behavior of the distribution. The $\delta$ parameter models the assumption that the density profile vanishes at an inner radius $r_\text{in}$, which has typically the same order of magnitude than the Schwarzschild radius of the BH at the center. Recent findings suggest that in order to assure energy conditions for the matter distribution we need to assume $r_\text{in}\geq\frac52 M_\text{BH}$ \cite{Shen2025}, where $M_\text{BH}$ is the BH mass (in absence of DM sources).

In this work we will study three examples of the generic density profile of Eq. (\ref{rho}), all of which reduce to known astrophysical density profiles of interest in the context of modeling DM halos and centers of elliptical galaxies. 

\subsubsection{Hernquist model}

The first of these models is associated with the parameters $(\alpha,\beta,\gamma,\delta)=(1,4,1,1)$, such that Eq. (\ref{eqm}) can be integrated in order to produce the following mass function 
\begin{equation}
    m_\text{H}(r)=M_\text{BH}+M_\text{H}\left(\frac{r-r_\text{in}}{r+a_0}\right)^2,
    \label{hernquist}
\end{equation}
where $M_\text{H}=\frac{2\pi a_0^3\rho_0}{1+r_\text{in}/a_0}$ is the total mass of the distribution. This density profile matches the famous Hernquist model \cite{Hernquist1990} in scales where the BH Schwarzschild radius $R_\text{S}=2M_\text{BH}$ can be neglected.

\subsubsection{NFW model}

The second model of interest is identified with parameters $(\alpha,\beta,\gamma,\delta)=(1,3,1,1)$, and the associated mass function is
\begin{equation}
    m_\text{NFW}(r)=M_\text{BH}+M_\text{NFW}\left[\log\left(\frac{r+a_0}{r_\text{in}+a_0}\right)-\frac{r-r_\text{in}}{r+a_0}\right],
\end{equation}
where $M_\text{NFW}=4\pi a_0^3\rho_0$. This density model matches the NFW model \cite{Navarro1997} at scales much larger than $R_\text{S}$. Observe that, despite the divergent character of the NFW mass function, the associated spacetime solution is still asymptotically flat, since $m_\text{NFW}(r)/r\to 0$ as $r\to\infty$. The NFW model is a successful analytical description of DM halos around galactic centers \cite{Navarro1997,Bertone2010}.

\subsubsection{Jaffe model}

The third model is identified with parameters $(\alpha,\beta,\gamma,\delta)=(1,4,2,1)$ and the associated mass function is
\begin{equation}
    m_\text{J}(r)=M_\text{BH}+\bar{M}_\text{J}\left[\frac{r-r_\text{in}}{r+a_0}+\frac{r_\text{in}}{a_0}\log\left(\frac{1+a_0/r}{1+a_0/r_\text{in}}\right)\right],
    \label{jaffe}
\end{equation}
where $\bar{M}_\text{J}$ is defined by
\begin{equation}
    \tilde{M_\text{J}}=\frac{M_\text{J}}{1-(r_\text{in}/a_0)\log(1+a_0/r_\text{in})}=4\pi a_0^3\rho_0,
\end{equation}
and $M_\text{J}$ is the total mass of the distribution. Observe that since $r_\text{in}\ll a_0$, these two masses are approximately the same. This density profile matches the Jaffe model \cite{Jaffe1983} at scales much larger than $R_\text{S}$. The Hernquist and the Jaffe profiles are successful analytical models for elliptical galaxies and bulges of disk galaxies \cite{Hernquist1990,Jaffe1983}.

Since the tangential pressure $q(r)$ is given in terms of the mass function $m(r)$ and the energy density $\rho(r)$, as shown in Eq. (\ref{eqq}), once we establish a density profile of matter, the tangential pressure profile is also determined. Due to the structure of Eq. (\ref{eqq}), if $r_\text{in}$ assumes any value larger than $R_S$, the pressure profile vanishes at this location. However, if $r_\text{in}=R_\text{S}$ then the pressure profile assumes a constant value at $r_\text{in}$. A simple calculation shows that those values are approximately equal to 
\begin{align}
    q_\text{H}(R_\text{S})M_\text{BH}^2&\sim\frac{1}{16\pi}\frac{M_\text{H}M_\text{BH}}{a_0^2},\\
    q_\text{NFW}(R_\text{S})M_\text{BH}^2&\sim\frac{1}{32\pi}\frac{M_\text{NFW}M_\text{BH}}{a_0^2},\\
    q_\text{J}(R_\text{S})M_\text{BH}^2&\sim\frac{1}{64\pi}\frac{M_\text{J}}{a_0}.
\end{align}
The above values serve also as upper bounds for the profile peak in the $r_\text{in}>R_\text{S}$ case. As we will discuss in the next section, for the astrophysically relevant values of the parameters $M_\text{BH},M$ and $a_0$, these upper bounds restrict the pressure profiles to be small in magnitude.

\section{Geodesic structure} \label{S:IV}

Consider the Lagrangian density of the orbital motion $\mathcal{L}=\frac12g_{\mu\nu}u^\mu u^\nu$, which using Eq. (\ref{metric}) can be written as
\begin{equation}
    2\mathcal{L}=-f(r)\dot{t}^2+\frac{\dot{r}^2}{h(r)}+r^2\dot{\theta}^2+r^2\sin^2\theta \ \dot{\varphi}^2=-\kappa,
\end{equation}
where $\kappa=0,1$ for null and timelike  geodesics, respectively, and the overdot stands for differentiation with respect to the affine parameter $\tau$ (the proper time for timelike geodesics). Since we are describing a spherically symmetric spacetime we can restrict the geodesic motion to the equatorial plane ($\theta=\pi/2$) without any loss of generality. Under this condition, from this Lagrangian density we can determine the canonical momenta conjugate to each of the remaining spacetime coordinates. In particular, we observe that the assumption of staticity and spherical symmetry implies the following conserved quantities
\begin{align}
    p_t&=-f(r)\dot{t}=-E,\label{energy}\\
    p_\varphi&=r^2\dot{\varphi}=L\label{angular},
\end{align}
where $E$ is the energy and $L$ is the angular momentum of the orbit.
Therefore we can effectively describe this orbital motion as an one-dimensional system in a two-dimensional phase space $(r,p_r)$. But before we do that it is useful to write the Lagrangian density of this one-dimensional motion incorporating the conserved quantities found above as
\begin{equation}
    2\mathcal{L}=\frac{\dot{r}^2}{h(r)}+\frac{L^2}{r^2}-\frac{E^2}{f(r)}=-\kappa,
    \label{lagrangian}
\end{equation}
which can be put in the form
\begin{equation}
    \frac{f(r)}{h(r)}\dot{r}^2+V_\text{eff}(r)=E^2,
    \label{geodesic}
\end{equation}
where we have introduced the effective potential for the one-dimensional system
\begin{equation} \label{eq:effpot}
    V_\text{eff}(r)=f(r)\left(\kappa+\frac{L^2}{r^2}\right).
\end{equation}
Note that Eq. (\ref{geodesic}) determine the radial motion of both null ($\kappa=0$) and timelike ($\kappa=1$) geodesics.

For later use, from the Lagrangian of Eq. (\ref{lagrangian}) we can determine the Hamiltonian of the system in the usual way
\begin{equation}
    2\mathcal{H}=2(p_\mu \dot{x}^\mu -\mathcal{L})=h(r)p_r^2+U_\text{eff}(r)=-\kappa,
    \label{hamiltonian}
\end{equation}
where we introduced another effective potential $U_\text{eff}(r)$ defined by  
\begin{equation}
    U_\text{eff}(r)=\frac{L^2}{r^2}-\frac{E^2}{f(r)}\label{potential2}
\end{equation}
purely for practical purposes.

\subsection{Circular geodesics}

Similarly to the Newtonian two-body problem, stationary points of the potential $V_\text{eff}(r)$ are associated with circular geodesics. Therefore we can look for circular orbits by imposing the following two conditions
\begin{equation}
    \dot{r}=0\quad\text{and}\quad V_\text{eff}'(r)=0.
\end{equation}
The first condition, together with Eq. (\ref{geodesic}), implies that the energy and the angular momentum of circular orbits must satisfy the relation
\begin{equation}
    E^2_c=f(r_c)\left(\kappa+\frac{L_c^2}{r_c^2}\right).
    \label{4.18}
\end{equation}
where $r_c$ is the circular orbit radius.

In order to impose the second condition for circular orbits we calculate the derivative of $V_\text{eff}(r)$ and search for stationary points. Recalling Eq. (\ref{eqf}) from the field equations of the Einstein cluster, we arrive at
\begin{equation}
    V'_\text{eff}(r_c)=\frac{2f(r_c)}{r^3(r_c-2m(r_c))}\left[\kappa r_c^2 m(r_c)-L^2(r_c-3m(r_c))\right]=0.
\end{equation}
Considering that the first factor cannot vanish, the equation for circular geodesics is
\begin{equation}
    \kappa r_c^2 m(r_c) -L^2[r-3m(r_c)]=0.
    \label{circular}
\end{equation}
Therefore the value of the angular momentum that produces circular orbits is
\begin{equation}
    L_c=\left[\frac{\kappa r_c^2 m(r_c)}{r_c-3m(r_c)}\right]^{1/2}
    \label{angularc}.
\end{equation}
Now using Eq. (\ref{4.18}) we conclude that the energy for circular orbits reads as\footnote{While this equation is not strictly valid for null geodesics, we prefer to leave the $\kappa$ in here.}
\begin{equation} 
    E_c=\left[\kappa f(r_c)\frac{r_c-2m(r_c)}{r_c-3m(r_c)}\right]^{1/2}.
    \label{energyc}
\end{equation}

From Eq. (\ref{circular}) we can set $\kappa=0$ in order to search for null circular geodesics. We arrive at the equation
\begin{equation}
    r_{\rm LR}-3m(r_{\rm LR})=0,
    \label{lightring}
\end{equation}
where we have denoted $r_{\rm LR}$ for the LR position. Notice that, although this equation is similar to the Schwarzschild case, in here $m$ \textit{is not} a constant, as it describes a mass function. In other words, for a given mass function $m(r)$, if there is a solution to the above equation at some radii, then at these locations there will be a LR, i.e. a null circular geodesic. Alternatively, we can set $\kappa=1$ and search for timelike circular geodesics using
\begin{equation}
     r_c^2m(r_c)-L^2[r_c-3m(r_c)]=0.
     \label{timelike}
\end{equation}
which are also of interest from the point of view of phenomenological applications.

\subsubsection{Angular velocity}

Observe that both expressions for $L_c$ and $E_c$, given by Eqs. (\ref{angularc}) and (\ref{energyc}), are undefined for the LR case. Nevertheless, the impact parameter $b_c=L_c/E_c$ is perfectly well behaved. In a similar fashion, we can define the coordinate angular velocity  associated with each orbit as \cite{Cardoso2009}
\begin{equation}
     \Omega \equiv \frac{\dot{\varphi}}{\dot{t}}=\frac{L}{E}\frac{f(r_c)}{r_c^2} \  ,
     \label{angularvelocity}
\end{equation}
where we used Eqs. (\ref{energy}) and (\ref{angular}) in the second equality. In order to determine $\Omega_c$ for circular orbits we substitute Eqs. (\ref{angularc}) and (\ref{energyc}) to arrive at
\begin{equation}
    \Omega_c=\frac{1}{r_c}\left[\frac{f(r_c)m(r_c)}{r_c-2m(r_c)}\right]^{1/2}.
    \label{angularvelocity2}
\end{equation}
Observe that the above equation is valid for both timelike and null circular geodesics. In the first case we only need to evaluate $\Omega_c$ at the location of the timelike geodesic, provided it exists. In the LR case, we can further use Eq. (\ref{lightring}) to simplify the expression to
\begin{equation} \label{angularvelocitylr}
\Omega_\text{LR}=\frac{\sqrt{f(r_\text{LR})}}{r_\text{LR}}.
\end{equation}
which is recognized as the angular velocity of unstable bound geodesics, a quantity of interest for the characterization of BH images.

\subsubsection{Stability analysis}

In order to determine the stability of these circular geodesics, we need to evaluate the second derivative of $V_\text{eff}(r)$ at these locations. If $V''_\text{eff}>0$ we have a stable geodesic, if $V''_\text{eff}<0$ we have an unstable geodesic, and if $V''_\text{eff}=0$ we have a marginally stable circular orbit (MSCO). Using the circular geodesic condition given by Eq. (\ref{circular}) we can express $V''_\text{eff}$ as
\begin{equation}
    V''_\text{eff}=\frac{2f}{r^4(r-2m)^2}\left[r^4m'(\kappa+L^2/r^2)+L^2(r-2m)(r-6m)\right].
    \label{second}
\end{equation}
For the null geodesic case, we set $\kappa=0$ and use Eq. (\ref{lightring}) to arrive at
\begin{equation}
    V''_\text{eff}=2fL^2\frac{r^2m'-3m^2}{r^4(r-2m)^2}.
\end{equation}
    Ignoring the always positive first factor and using Eq. (\ref{lightring}) again, we find that the LR is stable if
\begin{equation}
    3m'(r_\text{LR})>1.
    \label{lrstability}
\end{equation}
In standard BHs such a condition will never be satisfied and the LR will be unstable. Its fulfillment typically requires instead either exotic matter or the presence of UCOs.

In order to find the stability condition for the timelike case we go back to Eq. (\ref{second}), use Eq. (\ref{circular}) and set $\kappa=1$ to express $V''_\text{eff}$ as
\begin{equation}
    V''_\text{eff}=2f\frac{r^2m'+m(r-6m)}{r^2(r-2m)(r-3m)}.
\end{equation}
Since the first factor and the denominator are always positive for timelike circular geodesics, the condition for their stability is
\begin{equation}
    r^2m'(r)+m(r)[r-6m(r)]>0,
    \label{mscostability}
\end{equation}
and thus the condition for a MSCO is
\begin{equation}
    r^2m'(r)+m(r)[r-6m(r)]=0.
    \label{msco}
\end{equation}
Naturally, if there are multiple solutions to the above equation, the solution with the smaller radial coordinate would correspond to the so-called innermost stable circular orbit (ISCO). Observe that if there exists an interval of the radial coordinate in which the left-hand side of Eq. (\ref{msco}) is negative, then this would correspond to a region of unstable timelike circular orbits. This region would therefore not be populated with massive particles in a realistic description of the system, such as for the case of planar accretion disks.

\subsubsection{Lyapunov exponents}

Another relevant quantity for the characterization of the geodesic structure is the Lyapunov exponent (LE), which is a measure of the instability scale of a certain orbit. In the following construction of the LE we will follow closely the discussion of \cite{Cardoso2009,Deich2024}, though it can be alternatively derived from direct perturbation of the geodesic equation (\ref{geodesic}) itself. First we write the dynamical equations of the geodesic motion in the form
\begin{equation}
    \dot{X}^a=\omega^{ab}\partial_b\mathcal{H},
    \label{hamilton}
\end{equation}
where $X^a(t)$ is the set of coordinates in the associated phase space, $\omega^{ab}$ is the symplectic metric, $\mathcal{H}$ is the Hamiltonian of the system, and the overdot stands for differentiation with respect to some time parameterization. The latin indices represent phase space dimensions. To proceed, we linearize Eq. (\ref{hamilton}) considering a small deviation $\delta X^a(t)$ around some phase space trajectory $X^a(t)$. The deviation $\delta X^a(t)$ evolves according to
\begin{equation}
    \delta \dot{X}^a(t)={J^a}_b(t) \ \delta X^b(t),
    \label{deviation}
\end{equation}
where ${J^a}_b(t)$ is a symplectic Jacobian matrix defined by
\begin{equation}
    {J^a}_b(t)=\omega^{ac}\partial_c\partial_b\mathcal{H}(t).
    \label{jacobian}
\end{equation}
In the most general case, to define the LE associated with the trajectory $X^a(t)$ we would need to determine the eigenvalues of a global evolution operator defined by the time integral of ${J^a}_b(t)$ over the entire trajectory. In this specific case, since the dynamical equations of the system are autonomous, i.e., are not explicitly time-dependent, it is sufficient to determine the eigenvalues of ${J^a}_b$ itself.

Now, applying this machinery to the equatorial geodesic motion determined by the Hamiltonian of Eq. (\ref{hamiltonian}), ${J^a}_b$ is the following $2\times 2$ matrix 
\begin{equation}
{J^a}_b=
\begin{pmatrix}
    h'(r)p_r & h(r) \\
    -\frac12 U''_\text{eff}(r) & -h'(r)p_r
\end{pmatrix}.
\end{equation}
If we now particularize this result for a circular geodesic of radius $r_c$, where $p_r=0$, we get
\begin{equation}
\left.{J^a}_b\right|_\text{c}=
\begin{pmatrix}
    0 & h(r_c) \\
    -\frac12 U''_\text{eff}(r_c) & 0
\end{pmatrix},
\end{equation}
where the subscript indicate evaluation at $r_c$. As said above, the principal LE of the system is the larger eigenvalue of ${J^a}_b$, which is
\begin{equation}
    \lambda_p=\sqrt{-\frac{h_c}{2f_c}V''_\text{eff}(r_c)},
    \label{properexponent}
\end{equation}
where we have restored the original effective potential by the relation $V''_\text{eff}(r_c)=U''_\text{eff}(r_c)f_c$, and the subscript in $\lambda_p$ indicate that this is the LE with respect to the proper or affine parameterization. 

One should note that if one changes the time parameterization of the dynamical equations, the LE value will also change. In this sense, it is useful to determine the LE associated with the Schwarzschild time parameterization. To this end we make use of the fact that while the LE itself is not reparameterization-invariant, its product with a time interval measured in the same time parameterization is; in other words we have that $\lambda \delta t=\lambda_p \delta \tau,
$ where $\delta t$ is a time interval in Schwarzschild coordinates, $\lambda$ is the LE with respect to this time parameterization, and $\delta \tau$ is a proper time or affine parameter interval. Since we know that $\dot{t}=dt/d\tau=E/f(r)$ by Eq. (\ref{energy}), we arrive at
\begin{equation}
    \lambda=\sqrt{-\frac{h_cf_c}{2E^2}V''_\text{eff}(r_c)}.
    \label{schwexponent}
\end{equation}
In particular, the LE evaluated at the LR locations $r_\text{LR}$ can be calculated with the help of the field equation (\ref{eqf}) and Eq. (\ref{lightring}). The second derivative of $V_\text{eff}(r)$ evaluated at the LR radius $r_\text{LR}$ is
\begin{equation}
    V''_\text{eff}(r_\text{LR})=\frac{6L^2f_\text{LR}}{r_\text{LR}}(3m'_\text{LR}-1),
\end{equation}
from which, together with the identification of the angular velocity of Eq. (\ref{angularvelocitylr}), we conclude that the LR LE is given by
\begin{equation}
    \lambda_\text{LR}=\Omega_\text{LR}\sqrt{1-3m'_\text{LR}}.
    \label{lyapunov}
\end{equation}
Observe that this result is in agreement with the stability condition for the LRs given by Eq. (\ref{lrstability}), that is, a positive LE for an unstable LR and an imaginary LE for a stable LR. Note also that we can recover the result for the isolated Schwarzschild BH  simply by assuming a constant mass function. It is also worth pointing out that the above relation connects naturally two properties of features present in BH images, namely the shadow's size (i.e., the central region displaying a brightness deficit), associated to $\Omega_\text{LR}$, and the photon rings (the set of luminous rings surrounding the shadow), associated to the instability time-scale given by $\lambda_\text{LR}$, both entwined via the BH mass function.

\subsection{Low compactness configurations}

For the Schwarzschild spacetime, in which $m(r)=M_\text{BH}$, the expressions (\ref{lightring}) and (\ref{msco}) deliver the well known values for the existence of an unstable LR at $r=3M_\text{BH}$ and an ISCO at $r=6M_\text{BH}$, respectively. When we add a matter distribution around the BH with an inner radius at the horizon, i.e., $r_\text{in}=2M_\text{BH}$, the original LR and ISCO are now immersed in matter and their locations may change. Here we search for analytical approximations for the new LR and ISCO radii in the regime of low compactness of the BH surrounding matter. To make this condition precise we define the effective compactness of the distribution as
\begin{equation}
    C=\frac{M}{a_0},
    \label{compactness}
\end{equation}
where $M$ can be the total mass for the Hernquist and Jaffe density profiles and $4\pi\rho_0a_0^3$ for the NFW profile, since the latter does not have a convergent mass function. Therefore the regime we are interested in is characterized by
\begin{equation}
    \frac{M_\text{BH}}{a_0}\ll C\ll 1,
    \label{hierarchy}
\end{equation}
which is the one where the surrounding matter is sufficiently diffused for these models to be in agreement with observational data, such as accretion disk luminosity \cite{Boshkayev:2020kle,Heydari-Fard:2022xhr} or shadow observations \cite{Xavier:2023exm}.

\subsubsection{Analytical approximations for the LR location}

In order to estimate the LR location in this regime we need to search for a perturbative solution of Eq. (\ref{lightring}) for each of the density profile models. For the Hernquist model we find 
\begin{equation}
    r_\text{LR}^\text{Hern}\sim3M_\text{BH}\left(1+\frac{C_\text{H}M_\text{BH}}{a_0}\right).
    \label{lrhernquist}
\end{equation}
For the NFW model we find
\begin{equation}
    r_\text{LR}^\text{NFW}\sim3M_\text{BH}\left(1+\frac{C_\text{NFW}M_\text{BH}}{2a_0}\right).
    \label{lrnfw}
\end{equation}
Lastly, for the Jaffe model we find
\begin{equation}
    r_\text{LR}^\text{Jaffe}\sim3M_\text{BH}\left(1+\alpha_\text{J}C_\text{J}+\alpha_\text{J}C_\text{J}^2-\frac{C_\text{J}M_\text{BH}}{a_0}\right),
    \label{lrjaffe}    
\end{equation}
where $\alpha_\text{J}=1-2\log3/2\approx 0.2$ and the compactness quantities $C_\text{H},C_\text{NFW}$ and $C_\text{J}$ are defined according to Eq. (\ref{compactness}) using the associated total masses $M_\text{H},M_\text{NFW}$ and $M_\text{J}$ defined in section \ref{S:III}. We should note that in the above expressions we are omitting contributions of order $C^2 M_\text{BH}/a_0$ and smaller, in agreement with this low compactness approximation.

Although the LR position is direct to compute, there is an additional ingredient needed to compute the angular frequency and the LE, which is the redshift function $f(r)$. Later on we shall go back to this, describing an effective way to deal with $f(r)$ analytically.

\subsubsection{Analytical approximations for the ISCO location}

Now let us estimate the ISCO radius for each density profile models by searching for perturbative solutions of Eq. (\ref{msco}). For the Hernquist model, we find
\begin{equation}
    r_\text{ISCO}^\text{Hern}\sim 6M_\text{BH}\left(1-32\frac{C_\text{H}M_\text{BH}}{a_0}\right).
    \label{iscohern}
\end{equation}
For the NFW model, we find
\begin{equation}
    r_\text{ISCO}^\text{NFW}\sim6M_\text{BH}\left(1-16\frac{C_\text{NFW}M_\text{BH}}{a_0}\right).
    \label{isconfw}
\end{equation}
Lastly, for the Jaffe model, we find
\begin{equation}
    r_\text{ISCO}^\text{Jaffe}\sim 6M_\text{BH}\left(1-\beta_\text{J} C_\text{J}+\gamma_\text{J}C_\text{J}^2+32\frac{C_\text{J}M_\text{BH}}{a_0}\right),
    \label{iscojaffe}
\end{equation}
where $\beta_\text{J}=2\log(3)$ and $\gamma_\text{J}=16+2\beta_\text{J}$. We are again omitting contributions of order $C^2M_\text{BH}/a_0$ or smaller.

From the above analytical results, we conclude that when we add a matter distribution around the BH, such that the LR and the ISCO from the Schwarzschild solution are immersed in it, the former is pushed outwards while the latter is pulled inwards. This produces a smaller region between them, into which the accreting matter plunges, which might have a reflection on the observational appearance of these objects should the plasma there emit with sufficient intensity. On the other hand, observe that the Jaffe model results of Eqs. (\ref{lrjaffe}) and (\ref{iscojaffe}) stand out in comparison with the results from the other two models since it contains linear and quadratic contributions in $C_\text{J}$ only, which are substantially bigger than the $C_\text{J}M_\text{BH}/a_0$ contribution. This can be interpreted as an effect of the higher slope of the Jaffe density profile in the near zone $r\ll a_0$. Interestingly, the results for the Hernquist and the NFW models are strikingly similar. In fact, if we recall that $C_\text{H}\approx 2\pi\rho_0a_0^2=C_\text{NFW}/2$, the results are the same. This can be interpreted as a consequence of the equality of the slopes of both density profiles in the near zone $r\ll a_0$.

Since $r_\text{in}$ is the inner boundary of the matter distribution, and a free parameter of the model, one could argue that $r_\text{in}$ and $r_\text{ISCO}$ must coincide, whatever its value may be. If we follow this assumption, Eq. (\ref{msco}) must hold when evaluated at $r_\text{in}$, therefore vanishing the term proportional to $m'(r)$. Since $m(r_\text{in})=M_\text{BH}$, we conclude that $r_\text{ISCO}=6M_\text{BH}$, maintaining the result from the Schwarzschild solution \cite{Maeda2025}. 

\begin{figure*}[t!]
\centering
    \begin{subfigure}{0.48\textwidth}
        \includegraphics[width=\textwidth]{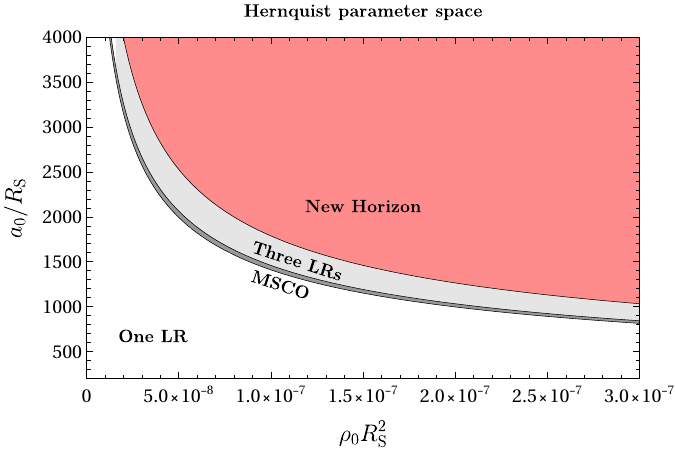}
        \label{fig:sub1}
    \end{subfigure}
    \begin{subfigure}{0.48\textwidth}
        \includegraphics[width=\textwidth]{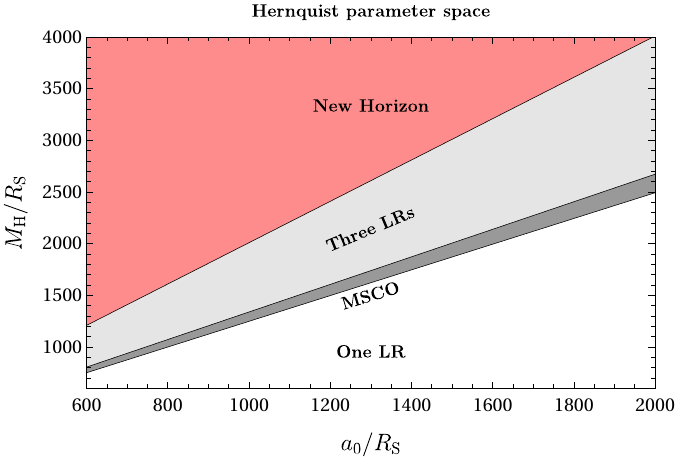}
        \label{fig:sub2}
    \end{subfigure}
    
    \vspace{\floatsep} 
    
    \begin{subfigure}{0.48\textwidth}
        \includegraphics[width=\textwidth]{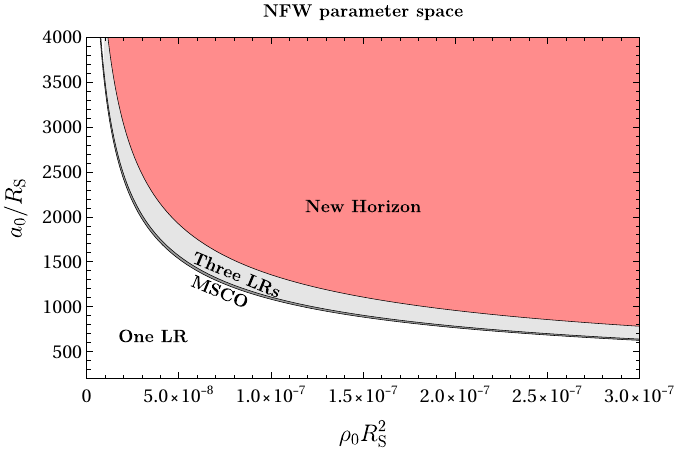}
        \label{fig:sub3}
    \end{subfigure}
    \begin{subfigure}{0.48\textwidth}
        \includegraphics[width=\textwidth]{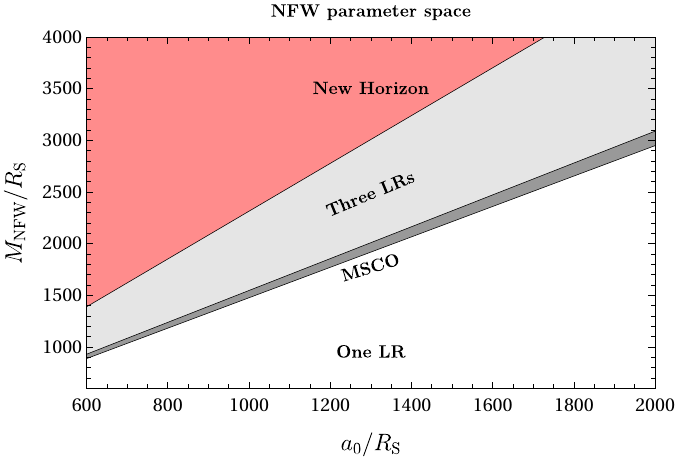}
        \label{fig:sub4}
    \end{subfigure}
    
    \vspace{\floatsep} 
    
    \begin{subfigure}{0.48\textwidth}
        \includegraphics[width=\textwidth]{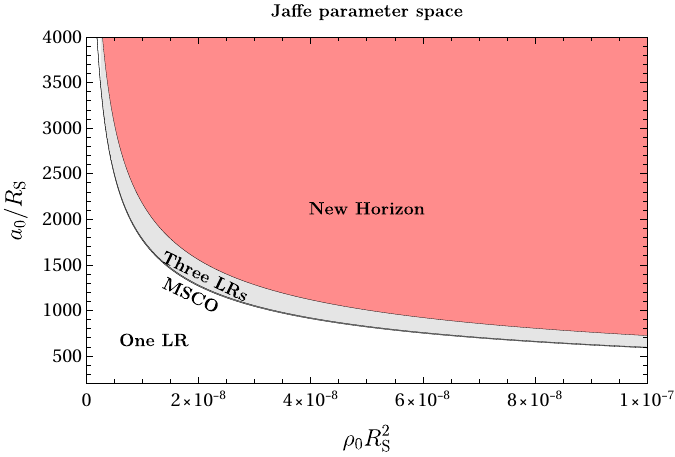}
        \label{fig:sub5}
    \end{subfigure}
    \begin{subfigure}{0.48\textwidth}
        \includegraphics[width=\textwidth]{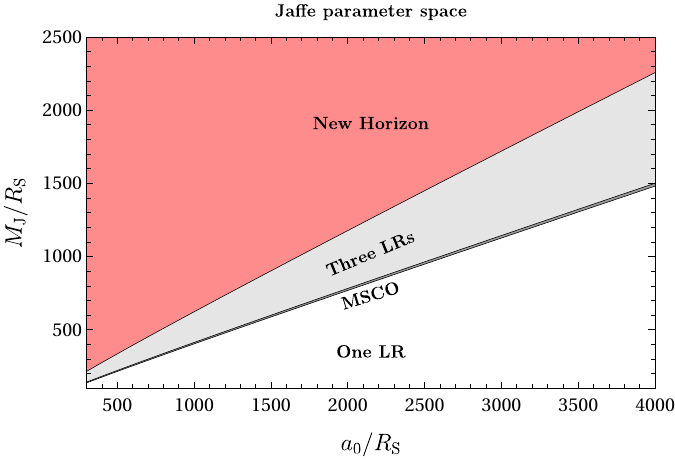}
        \label{fig:sub6}
    \end{subfigure}
    
    \caption{\justifying We consider the parameter space characterized by  $(\rho_0,a_0)$ in the first column and the parameter space characterized by  $(a_0,M)$ in the second column. The Hernquist, NFW and Jaffe models are presented in each row (from top to bottom). The red regions represent the formation of a new event horizon, the lighter gray region represents the formation of two additional LRs inside the matter distribution, and the darker gray regions represent the formation of a region of unstable timelike orbits also inside the matter distribution. In all plots we are considering $r_\text{in}=6M_\text{BH}$ and in the bifurcation plots we fixed $a_0=10^3 R_S$.}
    \label{fig1}
\end{figure*}

\begin{figure*}
\centering
    \begin{subfigure}{0.48\textwidth}
        \includegraphics[width=\textwidth]{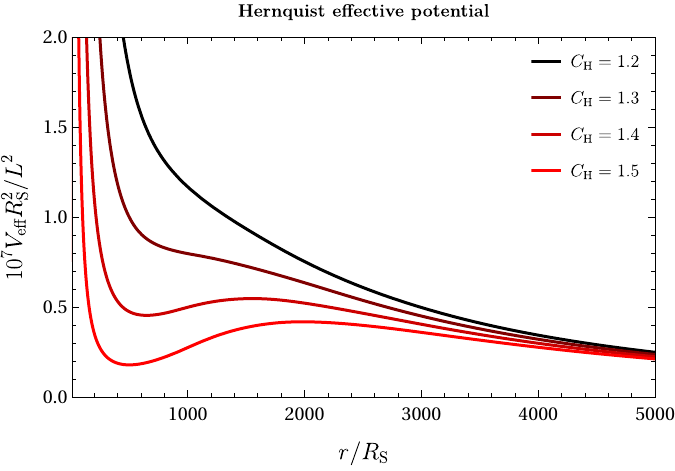}
    \end{subfigure}
    \begin{subfigure}{0.48\textwidth}
        \includegraphics[width=\textwidth]{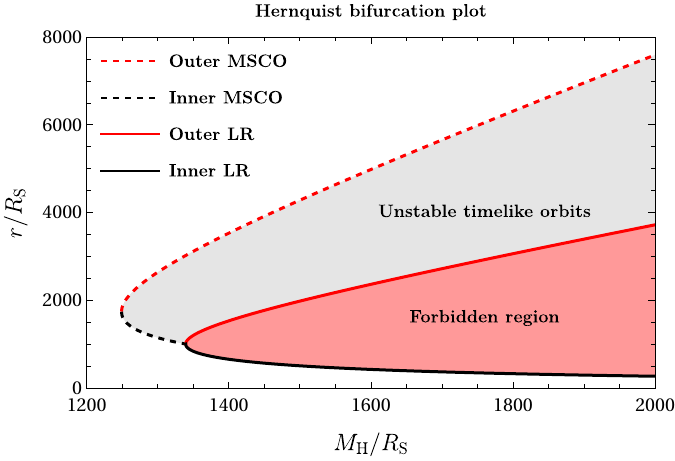}
    \end{subfigure}
    
    \vspace{\floatsep}
    
    \begin{subfigure}{0.48\textwidth}
        \includegraphics[width=\textwidth]{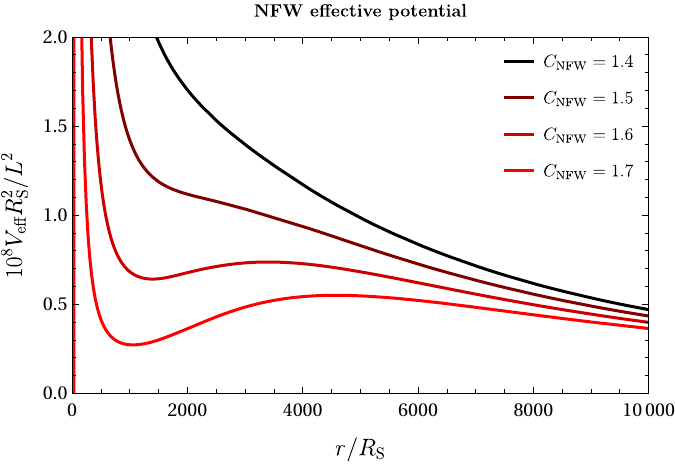}
    \end{subfigure}
    \begin{subfigure}{0.48\textwidth}
        \includegraphics[width=\textwidth]{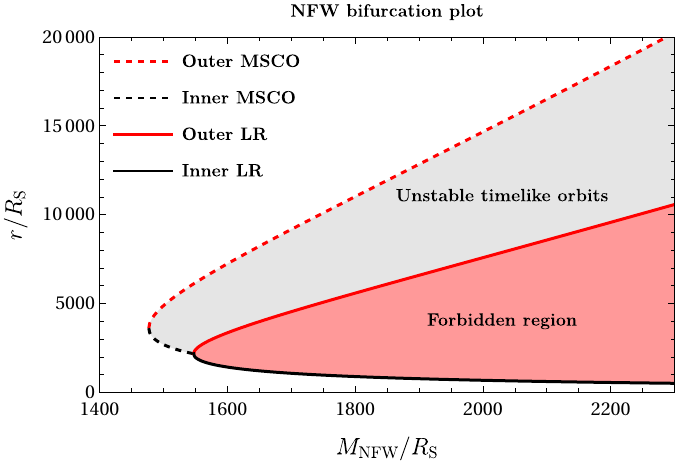}
    \end{subfigure}
    
    \vspace{\floatsep}
    
    \begin{subfigure}{0.48\textwidth}
        \includegraphics[width=\textwidth]{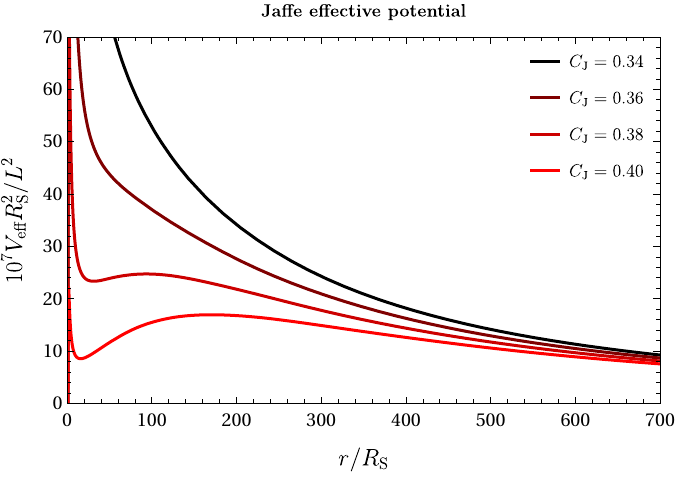}
    \end{subfigure}
    \begin{subfigure}{0.48\textwidth}
        \includegraphics[width=\textwidth]{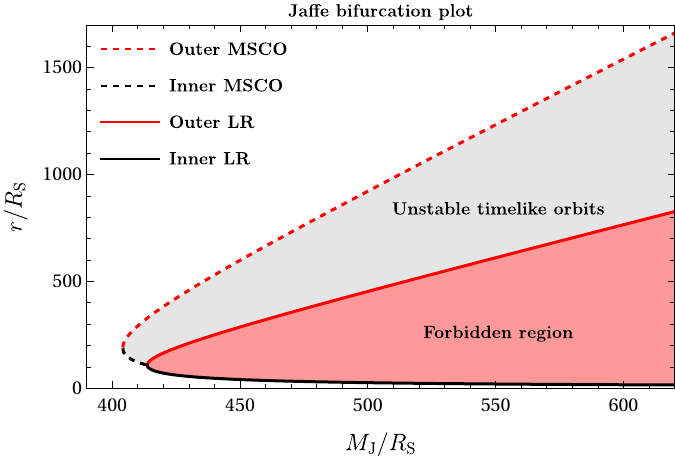}
    \end{subfigure}
    
    \caption{\justifying  In the first column of the above figure we display the effective potential of geodesic motion with increasing compactness for each model. In the second column we follow the changing radial coordinates of the additional LRs and MSCOs as the total mass increases, also highlighting relevant regions between these orbits. For the bifurcation plots we fixed $r_\text{in}=R_\text{S}$ and $a_0=10^3 R_\text{S}$. We highlight that in the white region we have stable orbits.}
    \label{figBIF}
\end{figure*}

\section{Ultracompact configurations} \label{S:V}

In this section, we investigate whether the three models we are considering exhibit additional LRs or MSCOs in comparison with the Schwarzschild solution. In other words, we will determine the region in the parameter space, defined by the pairs $(\rho_0,a_0)$ or $(M,a_0)$, that allows for solutions of Eqs. (\ref{lightring}) and (\ref{msco}) for each model. In analogy to compact stars, we shall refer to models with additional LRs as \textit{ultracompact} objects. An additional criterion for this analysis is to make sure that $r-2m(r)>0$ outside the event horizon \cite{Shen2025}, which prevents a coordinate singularity in Eqs. (\ref{eqf}) and (\ref{eqq}), but can also be thought of as a requirement of positive definiteness of the radial component of the metric expressed in Eq. (\ref{metric}). Physically, if an additional solution to $r-2m(r)=0$ exists, this would imply the existence of a new horizon at this location, suggesting the gravitational collapse of the matter inside this region. 

\subsubsection{Parameter space exploration}

In Fig. \ref{fig1}, we present the parameter space characterized by $(\rho_0,a_0)$ in the first column and the parameter space characterized by  $(a_0,M)$ in the second column. The darker gray region represents the formation of a patch of unstable circular timelike geodesics, defined by the space between the two emergent MSCOs. The lighter gray region represents the formation of two additional LRs inside the matter distribution. The red region represents the formation of a new horizon, and is therefore physically inaccessible. Observe that, in the first column, the parameter space of the three density profile models has an increasingly larger red region, and an increasingly smaller darker gray and lighter gray regions. In other words, we conclude that the Hernquist model has the largest parameter space region associated with the formation of additional LRs and MSCOs, while the Jaffe model has the smallest parameter space region.

In the second column, we also present parameter space plots, but characterized by the pair $(a_0,M)$. We found that the boundary curves between the regions of the first column are transformed to approximately straight lines in the second column, which means that there are specific values of compactness $C=M/a_0$ that characterize these boundaries, the corresponding numerical compactness values being displayed in Table \ref{table}. With this new parameter space characterization, we can see that even though the Jaffe model has the smallest region of additional structure emergence, the total mass values needed are smaller and therefore closer to the realistic regime of the model. In all of the parameter space plots of Fig. \ref{fig1}, we are considering $r_\text{in}=6M_\text{BH}$, while alternative values for $r_\text{in}$ such as $r_\text{in}=2M_\text{BH}$ were considered but with no substantial impact on the boundary curves. This is not surprising since the location in which these emergent structures appear is of the order of the length scale parameter $a_0$, which is always way larger than $r_\text{in}$.

\begin{table}[t!]
\centering
\begin{tabular}{c|c|c|c}
\hline\hline
$C$ & New MSCOs & New LRs & New horizon \\ 
\hline
Hernquist & 1.24 & 1.33 &  2.00 \\ 
\hline
NFW & 1.47 & 1.54 &  2.31 \\ 
\hline
Jaffe & 0.36 & 0.37 & 0.56 \\
\hline\hline
\end{tabular}
\caption{Compactness values that characterize the emergence of additional geodesic structures.}
\label{table}
\end{table}

\subsubsection{Effective potential and bifurcation plots}

By definition, the emergence of new LRs and MSCOs must be accompanied by new extrema points of the effective potential $V_\text{eff}(r)$ in Eq. (\ref{eq:effpot}). With the aim to check this, in the first column of Fig. \ref{figBIF} we show plots of $V_\text{eff}(r)$ with $\kappa=0$, \textit{i.e.}, the potential describing null geodesics, for all three density profile models. Since this potential depends on the redshift function $f(r)$, we used numerical solutions for the plotting. The plots confirm the emergence of two new extrema points, one local minimum and one local maximum, as the compactness $C$ exceeds the values of additional LRs formation, as given in Table \ref{table}.

In the second column of Fig. \ref{figBIF} we display the plots of the exact location of the additional LRs and MSCOs as the matter configuration becomes more compact. We fixed the length parameter to be $a_0=10^3 R_\text{S}$, and then we increased the total mass of the configuration $M$, which evidently increases the compactness $C=M/a_0$. Here we confirm that, as mentioned before, the pair of MSCOs forms at the same point and get further apart as $C$ increases. The same happens for the pair of additional LRs, but at slightly higher values of $C$. For this reason we will refer to these plots as bifurcation plots.

We found that in the physical region between the two additional MSCOs, the stability condition of Eq. (\ref{mscostability}) is no longer met, therefore the inner and outer MSCOs are the boundaries of a patch of unstable circular timelike geodesics, marked as the gray region in the bifurcation plots. On the other hand, we found that the inner and outer LRs are boundaries of a region where $r-3m(r)<0$. Therefore, the physical space between them, marked as red in the bifurcation plots, will not be populated with any circular geodesic. We can see the validity of such a statement from the fact that the angular momentum and energy of Eqs. (\ref{angularc}) and (\ref{energyc}) become imaginary if $r-3m(r)<0$. We conclude that the density profile model in which the additional MSCOs and LRs are closest to the central BH is the Jaffe model, while in the NFW model, this formation occurs further away. We have evaluated these bifurcation plots for other values of the parameter $a_0$, and the results were the same up to a rescaling of axes.

Due to the simplicity of the Hernquist mass function, given by Eq. (\ref{hernquist}), the coordinate singularity equation $r-2m(r)=0$ and the LR equation (\ref{lightring}) assume the form of cubic equations. Therefore, we can determine the compactness values that produce new real roots of these equations and, hence, the appearance of a new horizon or additional LRs. We found that the maximum compactness  $C^\text{NH}_\text{H}$, above which a new horizon appears, is
\begin{equation}
    C_\text{H}^\text{NH}=2+4(\eta-1)\frac{M_\text{BH}}{a_0}+\mathcal{O}\left(\frac{M_\text{BH}^2}{a_0^2}\right),
    \label{nhcompactness}
\end{equation}
and the compactness $C^\text{LR}_\text{H}$ above which the additional LRs appear is 
\begin{equation}
    C_\text{H}^\text{LR}=\frac43+\frac{4(2\eta-3)}{3}\frac{M_\text{BH}}{a_0}+\mathcal{O}\left(\frac{M_\text{BH}^2}{a_0^2}\right),
    \label{lrcompactness}
\end{equation}
where $r_\text{in}=\eta M_\text{BH}$ defines the parameter $\eta$.
If we choose $\eta=3$ we arrive at a simple expression for the location of the additional LRs $r_\text{LR}^\pm$, which is   
\begin{equation}
    \frac{r_\text{LR}^\pm}{a_0}=\frac12(3C_\text{H}-2)\pm\frac12\sqrt{3C_\text{H}(3C_\text{H}-12M_\text{BH}/a_0-4)}.
    \label{analyticallr}
\end{equation}
for the outer (unstable) and inner (stable) LRs, respectively. The above analytical results are in perfect agreement with the numerical results from before.

\section{Analytical and numerical solutions} \label{S:VI}

In this section we discuss the solutions for the redshift function $f(r)$, and possible approaches to find them for each density profile model. We also complete the characterization of the geodesic structure by calculating solution-dependent quantities such as the angular velocity $\Omega$ at the innermost LR and at the ISCO, and the LE, given by $\lambda$, at the LR. Of the three density profile models that we are considering, the only one that produces an exact solution of Eq. (\ref{eqf}) is the Hernquist model. The solution with $r_\text{in}=2M_\text{BH}$ was obtained in \cite{Cardoso2022} and a solution with a generic inner radius was obtained in \cite{Shen2025}.

\subsubsection{Post-Schwarzschild expansion of the field equation}

Unfortunately, the other two density profile models (NFW and Jaffe) do not produce known exact analytical solutions to the redshift field equation (\ref{eqf}). There is, however, a viable perturbative approach to search for approximate solutions. The strategy here is to explore the observation-based hierarchy of Eq. (\ref{hierarchy}) to perform a perturbative expansion of the field equation in the low compactness regime, \text{i.e.} $C\ll1$. First observe that we can rewrite Eq. (\ref{eqf}) as
\begin{equation}
    \frac{f'}{f}=\frac{1}{r}\left[-1+\left(1-\frac{2m}{ r}\right)^{-1}\right].
\end{equation}
Recalling that all the density models we are considering have a mass function of the form $m(r)=M_\text{BH}+\bar{m}(r)$, the above equation can be written as
\begin{equation}
    \frac{f'}{f}=-\frac{1}{r}+\frac{1}{r-2M_\text{BH}}\left(1-\frac{2\bar{m}}{r-2M_\text{BH}}\right)^{-1}.
    \label{fefbar}
\end{equation}
Therefore, we can interpret $\mathcal{P}(r)=2\bar{m}(r)/(r-2M_\text{BH})$ as a perturbative function upon the Schwarzschild spacetime, which is the solution to the unperturbed problem. Now, if we are able to show that $\mathcal{P}(r)$ is comparable with $C$ or smaller, then we can safely expand Eq. (\ref{fefbar}) in powers of $\mathcal{P}(r)$ and search for analytical solutions. If we expand Eq. (\ref{fefbar}) up to second order, which will be our chosen precision, we get
\begin{equation}
    \frac{f'}{f}=\frac{2M_\text{BH}}{r(r-2M_\text{BH})}+\frac{2\bar{m}}{(r-2M_\text{BH})^2}+\frac{4\bar{m}^2}{(r-2M_\text{BH})^3}+\mathcal{O}(\mathcal{P}^3).
    \label{PS}
\end{equation}
Since the zeroth-order term of the above expansion produces the Schwarzschild spacetime $f_\text{S}(r)=1-2M_\text{BH}/r$, we will refer to this expansion as post-Schwarzschild (PS) and the PS order will naturally be associated with the power of $\mathcal{P}(r)$ up to which Eq. (\ref{fefbar}) is expanded. Although unpopular, this nomenclature has been used in the literature in the context of computing geodesic motion around more general BHs \cite{Hackmann2008,Hackmann2012,García2015}.

Now, if we can assure that $\mathcal{P}(r)=2\bar{m}(r)/(r-2M_\text{BH})$ is of the order of the compactness $C$ of the models in the near zone, then we can safely expand the above field equation in powers of this quantity. In fact, one can show that for all three models the perturbative function $\mathcal{P}(r)$ is comparable with $C$ even if $r\sim a_0$. Observe that the assumption of asymptotic flatness of the spacetime fixes the behavior of the perturbative function $\mathcal{P}(r)$ at the far zone, \text{i.e.} $r\gg a_0$, such that $\lim\limits_{r/a_0 \to \infty} \mathcal{P}(r)=0$. 
Surprisingly, the three models present analytical solutions of Eq. (\ref{PS}), with an arbitrary inner radius $r_\text{in}$, and all of the solutions can be made asymptotically flat.

\subsubsection{Post-Schwarzschild solutions}

We present below the post-Schwarzschild solutions for the redshift function of all the density profile models. Even though there is a known exact analytical solution for the Hernquist model \cite{Cardoso2022,Shen2025}, we included the Hernquist PS solutions for completeness. Although we were unable to find exact analytical solutions, the expansions here can be analytically described up to arbitrary order. As long as we are in a perturbative regime, this should be enough to use the solutions studied here in a realistic physical scenario. For simplicity we fix the inner radius to be $r_\text{in}=2M_\text{BH}$. The 1PS solution for the Hernquist model is the simplest, given by
\begin{equation}
    f^{(1)}_\text{H}(r)=\left(1-\frac{2M_\text{BH}}{r}\right)\exp\left({-\frac{2M_\text{H}}{a_0+r}}\right).
\end{equation}
The NFW 1PS solution  is given by
\begin{align}
    f^{(1)}_\text{NFW}(r)&=\left(1-\frac{2M_\text{BH}}{r}\right)\left(\frac{r+a_0}{x}\right)^{-\Upsilon^{(1)}_\text{NFW}},\\
    \Upsilon^{(1)}_\text{NFW}(r)&=\frac{2M_\text{NFW}}{r-2M_\text{BH}},\quad x=a_0+2M_\text{BH}.
\end{align}
Lastly, the Jaffe 1PS solution is given by 
\begin{align}
    f^{(1)}_\text{J}(r)&=\left(1-\frac{2M_\text{BH}}{r}\right)\left(1+\frac{a_0}{2M_\text{BH}}\right)^{\Upsilon^{(1)}_\text{J}}\left(1+\frac{a_0}{r}\right)^{\Lambda^{(1)}_\text{J}},\\
    \Upsilon^{(1)}_\text{J}(r)&=\frac{4C_\text{J}M_\text{BH}}{r-2M_\text{BH}}, \quad \Lambda^{(1)}_\text{J}(r)=-\Upsilon^{(1)}_\text{J}(r)-2C_\text{J}.
\end{align}

Now we present the 2PS solutions for the Hernquist and NFW models. For the Hernquist model we have
\begin{align}
    f^{(2)}_\text{H}(r)&=\left(1-\frac{2M_\text{BH}}{r}\right)e^{\Upsilon^{(2)}_\text{H}},\nonumber\\
    \Upsilon^{(2)}_\text{H}(r)&=-\frac{2M_\text{H}}{r+a_0}\left[1+\frac{M_\text{H}(3r+a_0-4M_\text{BH})}{3(r+a_0)^2}\right]\label{HernquistPS2}.
\end{align}
For the NFW we have
\begin{align}
    f^{(2)}_\text{NFW}(r)&=\left(1-\frac{2M_\text{BH}}{r}\right)\left(\frac{r+a_0}{x}\right)^{\Upsilon^{(2)}_\text{NFW}}\exp \Lambda^{(2)}_\text{NFW},\nonumber\\
    \Upsilon^{(2)}_\text{NFW}(r)&=\frac{4M_\text{NFW}^2}{x^2}\Bigg\{1-\log\left(\frac{r-2M_\text{BH}}{x}\right)\nonumber\\
    &-\frac12\left[1+\frac{x^2}{(r-2M_\text{BH})^2}\right]\log\left(\frac{r+a_0}{x}\right)\nonumber\\
    &+\frac{\left(2x-x^2/2M_\text{NFW}-r-a_0\right)}{r-2M_\text{BH}}\Bigg\},\nonumber\\
    \Lambda^{(2)}_\text{NFW}(r)&=\frac{4M_\text{NFW}^2}{x^2}\Bigg\{\frac{x}{r+a_0}-2\text{Li}_2\left(\frac{2M_\text{BH}-r}{x}\right)\nonumber\\
    &-\text{Re}\left[\text{Li}_2\left(\frac{r+a_0}{x}\right)\right]\Bigg\},
\end{align}
where $\text{Li}_2(z)$ is the dilogarithm function. We note that the Jaffe 2PS solution exists in an analytical form but it is extremely lengthy and unpractical to be written here.

\subsubsection{Angular velocities and Lyapunov exponents}\label{sec:anaytical_sol}

With the approximate analytical solutions in hand, we can evaluate the above mentioned relevant observable quantities using Eqs. (\ref{angularvelocity2}), (\ref{angularvelocitylr}) and (\ref{lyapunov}), together with the analytical approximations of the location of the LR and the ISCO from the previous section. Since we want to investigate the frequencies of the innermost LR and the ISCO as they are immersed in the matter configuration, we set $r_\text{in}=2M_\text{BH}$. The LR angular velocity for the Hernquist model is given by
\begin{equation}
    M_\text{BH}\Omega_\text{LR}^\text{Hern}\sim\frac{1}{3\sqrt{3}}\left(1-C_\text{H}+\frac16C_\text{H}^2+3\frac{C_\text{H}M_\text{BH}}{a_0}\right).
\end{equation}
For the NFW model, we have
\begin{align}
    M_\text{BH}\Omega_\text{LR}^\text{NFW}\sim& \frac{1}{3\sqrt{3}}\Bigg(1-C_\text{NFW}+\frac{\omega_\text{NFW}}{6}C_\text{NFW}^2\nonumber\\
    &+\frac52\frac{C_\text{NFW}M_\text{BH}}{a_0}\Bigg),   
\end{align}
where $\omega_\text{NFW}=21-2\pi^2\approx 1.3$. For the Jaffe model, we have  
\begin{equation}
    M_\text{BH}\Omega_\text{LR}^\text{Jaffe}\sim\frac{1}{3\sqrt{3}}\left[1-\left(\log\frac{a_0}{M_\text{BH}}-\log \frac{27}{4}\right)C_\text{J}\right].
\end{equation}

We can also calculate the angular velocities at the ISCO for each model. For the Hernquist model, we have
\begin{equation}
    M_\text{BH}\Omega_\text{ISCO}^\text{Hern}\sim\frac{1}{6\sqrt{6}}\left(1-C_\text{H}+\frac16C_\text{H}^2+66\frac{C_\text{H}M_\text{BH}}{a_0}\right).    
\end{equation}
For the NFW model, we have
\begin{align}
    M_\text{BH}\Omega_\text{ISCO}^\text{NFW}\sim&\frac{1}{6\sqrt{6}}\Bigg(1-C_\text{NFW}+\frac{\omega_\text{NFW}}{6}C_\text{NFW}^2\nonumber\\
    &+34\frac{C_\text{NFW}M_\text{BH}}{a_0}\Bigg). 
\end{align}
For the Jaffe model, we have
\begin{equation}
    M_\text{BH}\Omega_\text{ISCO}^\text{Jaffe}\sim\frac{1}{6\sqrt{6}}\left[1-\left(\log\frac{a_0}{M_\text{BH}}-\log54-3\right)C_\text{J}\right].
\end{equation}
It should be noted that the second order contribution for both the LR and ISCO frequencies of the Jaffe model are computable but rather lengthy to display here.

Using Eq. (\ref{lyapunov}) together with the previous results, we can also make analytical approximations for the LE at the innermost LR. For the Hernquist case we have
\begin{equation}
    M_\text{BH}\lambda_\text{LR}^\text{Hern}\sim\frac{1}{3\sqrt{3}}\left(1-C_\text{H}+\frac16C_\text{H}^2\right).
\end{equation}
For the NFW case we find
\begin{align}
    M_\text{BH}\lambda_\text{LR}^\text{NFW}\sim&\frac{1}{3\sqrt{3}}\Bigg(1-C_\text{NFW}+\frac{\omega_\text{NFW}}{6}C_\text{H}^2\nonumber\\
    &+\frac{C_\text{NFW}M_\text{BH}}{a_0}\Bigg).
\end{align}
For the Jaffe case we get
\begin{equation}
    M_\text{BH}\lambda_\text{LR}^\text{Jaffe}\sim\frac{1}{3\sqrt{3}}\left[1-\left(\log\frac{a_0}{M_\text{BH}}-\log \frac{27}{4}+\frac12\right)C_\text{J}\right].
\end{equation}

As a confirmation of the perturbative approach, we calculated the above Hernquist model results from both the 2PS and the exact solutions. We found agreement up to the desired precision. The above analytical approximations show that when the innermost LR and the ISCO are immersed in the matter distribution the associated angular velocities are decreased in comparison with the vacuum case. Observe that the similarity between the first order contributions of the Hernquist and NFW models can be explained by the agreement of their behavior at the near zone, whereas the distinct near zone behavior of the Jaffe model produces a difference already at the first order contribution in all of the above quantities.

\subsubsection{Numerical solutions}

Now we will compare the approximate analytical solutions, obtained from the PS scheme discussed above, with the solutions obtained with numerical methods. In order to obtain numerical solutions to Eq. (\ref{eqf}), we need to establish a cutoff radius $r_\text{cutoff}$ at which the asymptotic flatness boundary condition must be imposed, \text{i.e.} $f(r_\text{cutoff})=1$. Since we want to explore both the near zone ($r\ll a_0$) and the far zone ($r\gg a_0$) regions of these solutions, we will chose $r_\text{cutoff}$ to be significantly larger than $a_0$.

\begin{figure*}[t!]
\centering
    \begin{subfigure}{0.48\textwidth}
        \includegraphics[width=\textwidth]{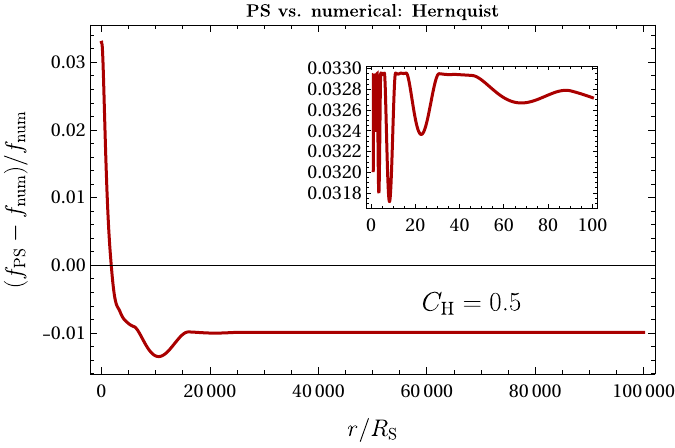}
    \end{subfigure}
    \begin{subfigure}{0.48\textwidth}
        \includegraphics[width=\textwidth]{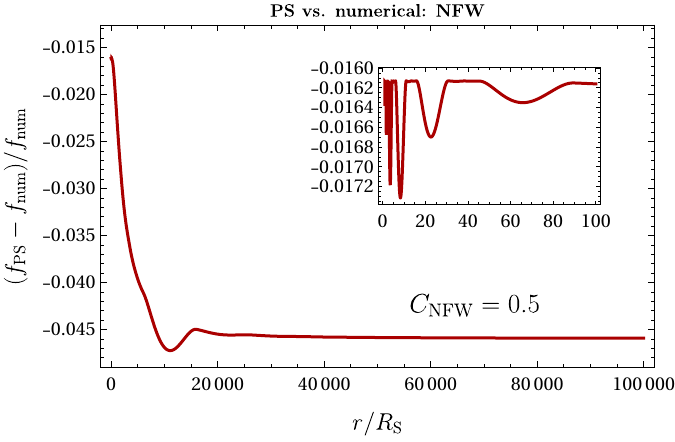}
    \end{subfigure}
        
    \begin{subfigure}{0.48\textwidth}
        \includegraphics[width=\textwidth]{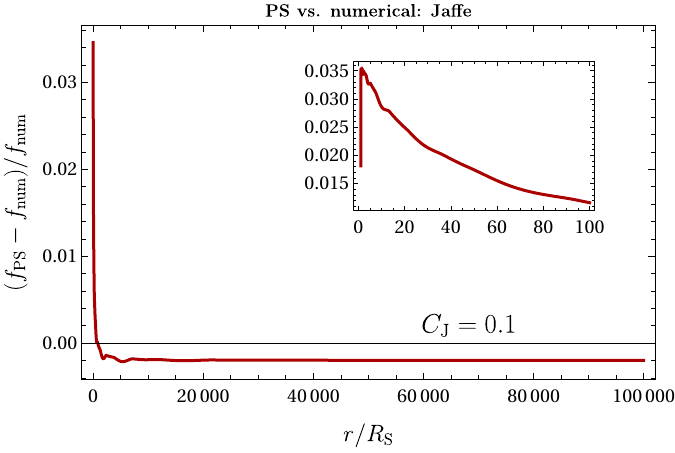}
    \end{subfigure}
        
    \caption{\justifying In this figure we show the relative error between the second order PS and the numerical solutions for all the models. We chose the length parameter to be $a_0=10^3R_\text{S}$, and the inner radius to be $r_\text{in}=R_\text{S}$. For the numerical solutions we chose $r_\text{cutoff}=10^5R_\text{S}$.}
    \label{figPSError}
\end{figure*}

The relative error $(f_\text{PS}-f_\text{num})/f_\text{num}$ between the 2PS and numerical solutions is presented in Fig. \ref{figPSError}. In these plots we chose the length parameter to be $a_0=10^3R_\text{S}$, and the inner radius to be $r_\text{in}=R_\text{S}$. For the numerical solutions we chose $r_\text{cutoff}=10^5R_\text{S}$. For the Hernquist and NFW models we used a compactness value of $C=0.5$, and for the Jaffe model a compactness of $C_\text{J}=0.1$, as displayed in the plots. We found that the 2PS approximate solutions are accurate both in the near zone and in the far zone regions.
We see that the error never exceeds $5\%$ for all the density profile models, even though the compactness values used are relatively high. This result substantiates the analytical approximations for the geodesic quantities found in section \ref{sec:anaytical_sol}.

\section{Ringdown analysis: parametric deviation, additional modes and parasitic BHs} \label{S:VII}

From the above features related to LRs we see that, depending on the parameter space we explore, we can have one or three LRs. It is well known that LRs and their associated quantities can also be tightly attached to the natural oscillation frequency of the spacetime, i.e. the quasinormal modes (QNM),  in the eikonal limit~\cite{Cardoso2009,Konoplya:2017wot,Pedrotti:2025idg}. Additionally, stable LRs are linked to long-lived, trapped modes, which are conjectured to lead to instabilities when nonlinear perturbations are considered~\cite{Cardoso:2014sna,Keir:2014oka} (see also \cite{Cunha:2022gde}). In this Section, we evolve initial data of test fields to elucidate the effect of additional LRs on the ringdown of a BH surrounded by matter.

For simplicity, we shall focus on scalar waves $\Psi$ propagating in the BH-DM spacetime, that is
\begin{equation}
    \frac{1}{\sqrt{-g}}\partial_\mu(\sqrt{-g}\partial^\mu\Psi)=0.
\end{equation}
By using a harmonic decomposition of the scalar field into $\Psi=\psi(t,r)Y_{\ell m}(\theta,\phi)/r$, with $Y_{\ell m}(\theta,\phi)$ being the spherical harmonics, we obtain the following wave equation
\begin{equation}\label{perteq}
\frac{\partial^2\psi(t,r_*)}{\partial t^2}-\frac{\partial^2\psi(t,r_*)}{\partial r_*{^2}}+V_{\psi}\psi(t,r_*)=0,
\end{equation}
with the {\it tortoise} coordinate $r_*$ being defined by $dr_*/dr=1/\sqrt{fh}$, and the effective potential being
\begin{equation}\label{eq:wave_psi}
    V_{\psi}=f\left[\frac{\ell(\ell+1)}{r^2}+\frac{2m}{r^{3}}-\frac{m'}{r^2}\right],
\end{equation}
where we have used the field equations to eliminate the derivatives of $f$. This potential fulfills the conditions stated in \cite{Glampedakis:2019dqh} for the correspondence between QNMs and BH imaging in the eikonal limit.

We transform the wave equation \eqref{eq:wave_psi} to null coordinates, using $du=dt-dr_{*}$ and $dv=dt+dr_{*}$, which leads to the wave equation
\begin{equation}
    4\frac{\partial^{2}\psi(u,v)}{\partial v\partial u}+V_{\psi}(r)\psi(u,v)=0
\end{equation}
and evolve it in a $u-v$ grid, using the Gundlach-Price-Pullin scheme as~\cite{Gundlach:1993tp}
\begin{equation}
    \psi(N)=\psi(W)+\psi(E)-\psi(S)-h^{2}V_{\psi}(S)\frac{\psi(W)+\psi(E)}{8}+\mathcal{O}(h^{4}),
\end{equation}
subject to the following initial conditions
\begin{equation}
\psi(0,v)=A_{0}\text{exp}[-(v-v_{0})^{2}/2\sigma], \,\psi(u,0)=0,  
\end{equation}
with $(A_{0},v_{0},\sigma)=(1,10,1)$, where $A_{0}$, $v_{0}$, and $\sigma$ are the Gaussian's height, location, and width, respectively. To evolve the initial conditions, we numerically integrate the Einstein field equation~\eqref{eqf} to obtain the lapse function, employing a root-finding procedure at each step to determine $r(r_*)$. After the evolution, we extract the wave signal at some far region, typically $r_*=200 a_{0}$. We used $h=10^{-1}$ for the grid spacing, higher resolution presented similar qualitative results for the ringdown profile, and we did not require resolution to see the gravitational tail, as this is not the focus of the present work. Quantitatively, the errors were not relevant for the overall discussion of the modes nature, which were insensitive to decreasing $h$ for the numbers we discuss. The extracted signal is then analyzed, searching for the oscillation frequencies and damped behavior.

The configuration presents two distinct length scales, one associated with the BH  event horizon, $R_\text{S}$, and the other with the radius of the DM halo $a_{0}$. In the case of a low compactness regime, and to compare between different configurations and the vacuum case, we evolved using the event horizon scale. For the high compactness regime, we used the DM halo radius scale, since it proved useful to parametrize the quantities in terms of the DM radius in this case.

\begin{figure*}[t!]
    \includegraphics[width=.47\textwidth]{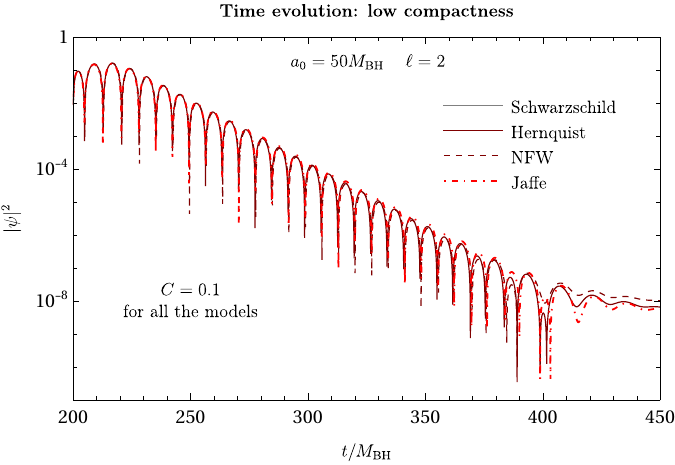}
    \includegraphics[width=.47\textwidth]{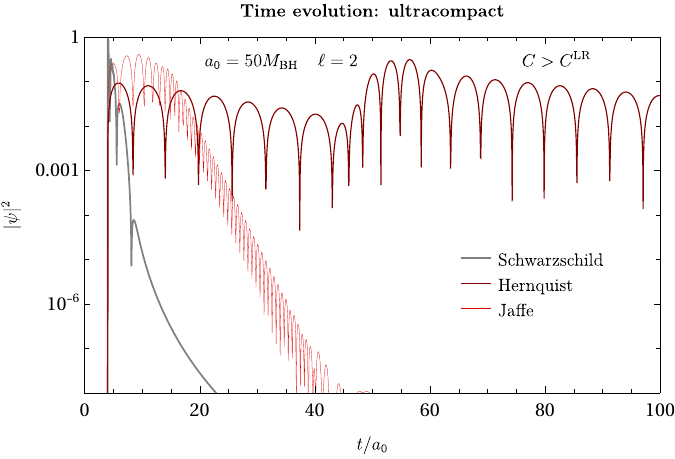}
     \caption{\justifying  We have the time evolution comparison between the cases with and without the DM halo. In the left panel, we have low compactness regime, and in the right panel, we have compactness $C>C^\text{LR}$ for each case.}
     \label{FIG:TEComparison}
     \end{figure*}

It is worth to recall the connection between the ringdown frequencies and the null geodesic quantities. Considering the eikonal approximation, we have that the fundamental QNM related to maximum of the potential can be written as~\cite{Mashhoon:1982im,Schutz1985,Pedrotti:2025idg}
\begin{equation}\label{eq:mode}
\omega_{\rm QNM}\approx \ell \Omega_{\rm LR} -i |\lambda_{\rm LR}|/2.
\end{equation}
where $\Omega_\text{LR}$ and $\lambda_\text{LR}$ are given in Eqs. (\ref{angularvelocitylr}) and (\ref{schwexponent}), respectively. Naturally, when low compactness scenarios are considered, inserting the corresponding equations for these quantities discussed in the above section for the Hernquist, NFW, and Jaffe models, leads to parametric deviation of the Schwarzschild modes (see Sec. \ref{sec:anaytical_sol}). As such, we expect that the ringdown in such scenarios to be qualitatively very similar to the Schwarzschild case, that is a damped sinusoidal behavior, though with some quantitative differences.

In Fig.~\ref{FIG:TEComparison}, we show some representative ringdowns. In the low compactness regime (left panel of Fig.~\ref{FIG:TEComparison}), we can see that the field evolves in a very similar fashion to the Schwarzschild BH, shown together for comparison. This is a direct result of the parametric deviation of the potential, which is known to smoothly deviate the ringdown from the Schwarzschild counterpart. We also show the extracted frequencies and their behavior with respect to low compactness in Table ~\ref{tab:mode_comp} (the row for $C=0.1$). For comparison, Table \ref{tab:mode_comp} also displays the QNM frequencies computed via the well known third-order WKB method \cite{IyerWill1987,Iyer1987}. The relative error between the two methods never exceed 5\% and the results are in agreement with the deviations expected from the geodesic limit explained above.

In the right panel of Fig.~\ref{FIG:TEComparison}, we present high-compactness scenarios, being high enough to have an additional pair of stable-unstable LRs (in agreement with the arguments of \cite{Cunha:2017qtt} that for UCOs such structures always come in pairs), i.e., $C>C^\text{LR}$\footnote{\label{fotnoteNFW}We do not include the results for the NFW case in this plot, because it presents a redshift problem which could not be solved with the numerical methods used in this work.}. The additional local maximum in the potential can bring a new spectrum which would be similar to the one given by Eq.~\eqref{eq:mode}. Apart from that, the stable LR itself has modes related to it, that are trapped in between the two maxima, similarly to a potential well in standard quantum mechanics. These types of trapped modes have been studied in ultracompact relativistic stars~\cite{Kokkotas1994,Andersson:1995ez,Kokkotas:1999bd}, and their geodesic correspondence can also be viewed within a WKB approach~\cite{Gurvitz1988,Cardoso:2014sna,Guo:2021bcw}. To highlight the difference between these and the modes related to the potential peak, we shall refer to them as trapped states (TS). The fundamental mode frequency related to the stable LR can be written as (see \cite{Cardoso:2014sna,Guo:2021enm} for an in-depth discussion)
\begin{equation}
    \omega_\text{TS}\approx\Omega_{\rm SLR}\ell -i b e^{-c\ell},
\end{equation}
with $\Omega_{\rm  SLR}$ being the frequency of the stable LR, and $b$ and $c$ being positive constants. This means that at the eikonal limit, these modes live extremely long, as one might expect due to the relation with stable orbital configurations.

\begin{figure*}
    \includegraphics[width=.47\textwidth]{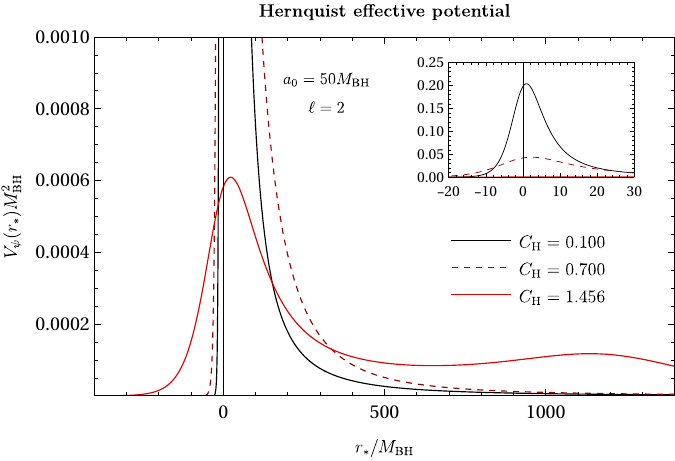}
    \includegraphics[width=.49\textwidth]{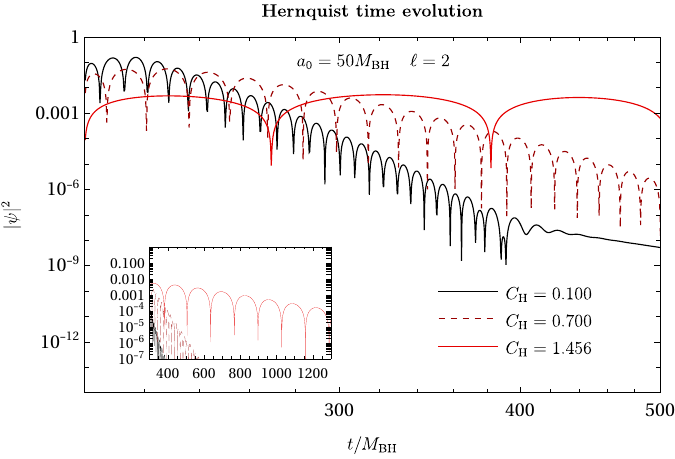} 
     \caption{\justifying We have the potential (left panel) and the time evolution (right panel) for a BH with a DM halo for a Hernquist density profile. Increasing the DM halo compactness leads to significant modifications in the ringdown oscillation frequency and decay time.}
     \label{FIG:TEHern_Vara0}
     \end{figure*}

\begin{figure*}
    \includegraphics[width=.47\textwidth]{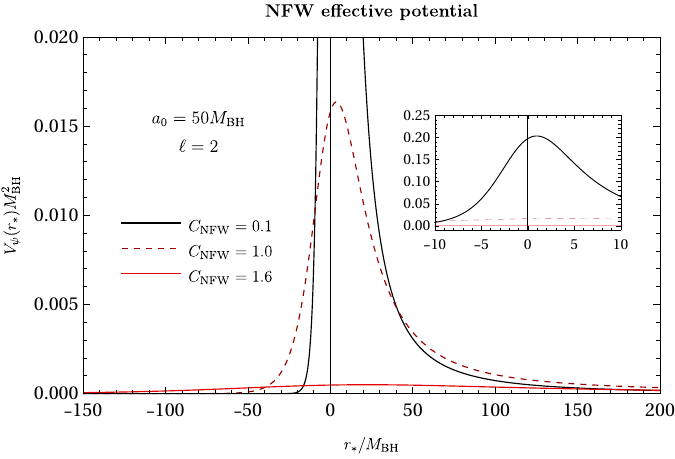}
    \includegraphics[width=.49\textwidth]{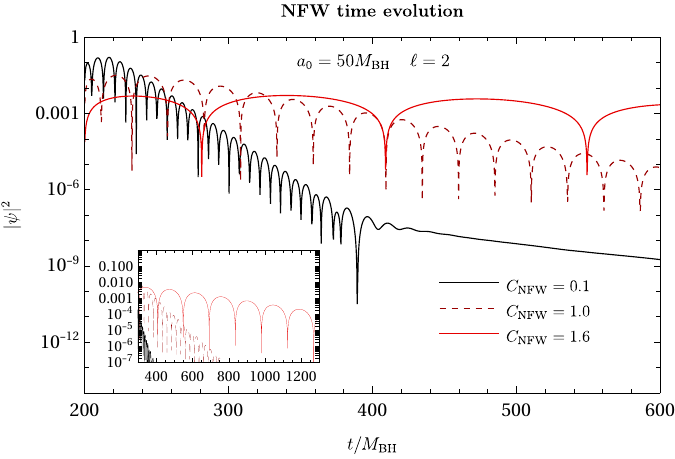} 
     \caption{\justifying We have the potential (left panel) and the time evolution (right panel) for a BH with a DM halo for a NFW density profile. Increasing the DM halo compactness leads to significant modifications in the ringdown oscillation frequency and decay time.}
     \label{FIG:TENFW}
     \end{figure*}

\begin{figure*}
    \includegraphics[width=.47\textwidth]{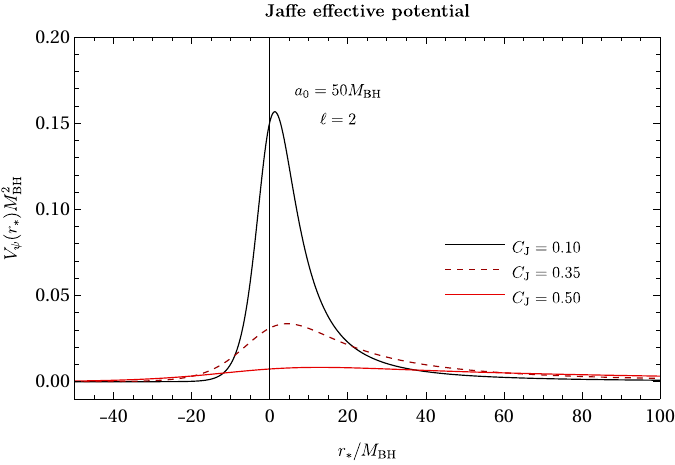}
    \includegraphics[width=.49\textwidth]{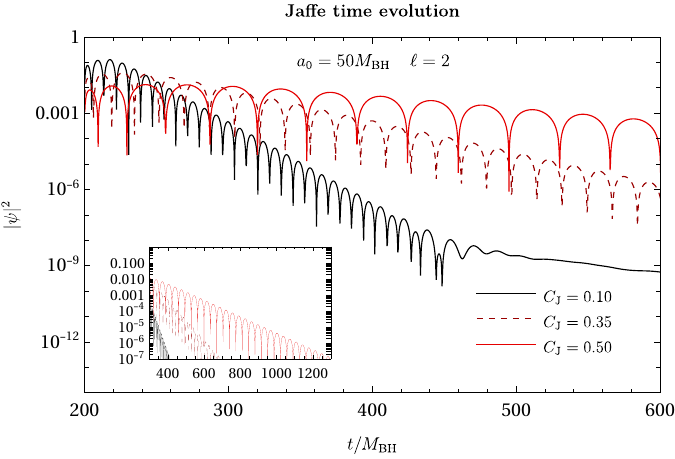} 
     \caption{\justifying We have the potential (left panel) and the time evolution (right panel) for a BH with a DM halo for a Jaffe density profile. Increasing the DM halo compactness leads to significant modifications in the ringdown oscillation frequency and decay time.}
     \label{FIG:TEJaffe}
\end{figure*}

One existing feature related to the trapped modes in ultracompact stars is the presence of echoes on the ringdown~\cite{Ferrari:2000sr,Cardoso:2016rao,Cardoso:2016oxy}, where the ringdown signal is followed by tunneled wave-packets after successive reflections from the potential barrier. A spectral analysis of these types of signal clearly shows the correspondence with the TS described above. Therefore, it is natural to expect, in some regime, to have some echo-like feature when one consider BHs surrounded by matter in the high compactness scenario. Notice, however, that a clear echo picture will only be visible if the travel time scale between the maxima is compatible with the ringdown scale in the signal~\cite{Cardoso:2019rvt}. 

\begin{table*}
    \centering
    \begin{tabular}{c|cc|cc|cc}
    \hline\hline
                & \multicolumn{2}{c|}{Hernquist} & \multicolumn{2}{c|}{NFW} & \multicolumn{2}{c}{Jaffe} \\
                & Prony & WKB & Prony & WKB & Prony & WKB \\
    \hline
    Solid black ($C=0.1$) 
        & $0.451 - 0.089 i$ & $0.451 - 0.087 i$ 
        & $0.453 - 0.089 i$ & $0.438 - 0.087 i$
        & $0.393 - 0.075 i$ & $0.385 - 0.073 i$ \\
    \hline
     Dashed red ($C\lesssim C^\text{LR}$)  
        & $0.206 - 0.040 i$ & $0.202 - 0.039 i$ 
        & $0.126 - 0.024 i$ & $0.124 - 0.024 i$ 
        & $0.181 - 0.029 i$ & $0.179 - 0.029 i$ \\
    \hline
    Solid red ($C>C^\text{LR}$)  
        & $0.024 - 0.004 i$ & $0.024 - 0.004 i$ 
        & $0.022 - 0.004 i$ & $0.022 - 0.004 i$
        & $0.090 - 0.012 i$ & $0.089 - 0.012 i$ \\
    \hline\hline
    \end{tabular}
    \caption{\justifying The quasinormal frequencies, $\omega M_\text{BH}$, extracted using the prony and third-order WKB methods from the ringdown presented in Figs.~\ref{FIG:TEHern_Vara0}, \ref{FIG:TENFW}, and \ref{FIG:TEJaffe}, for the three DM profiles.}  
    \label{tab:mode_comp}
\end{table*}

In Figs.~\ref{FIG:TEHern_Vara0}, \ref{FIG:TENFW} and \ref{FIG:TEJaffe}, we show the ringdown and the potential for different compactness regimes, for each DM profile, while the corresponding QNM frequencies, $\omega M_\text{BH}$, are displayed in Table \ref{tab:mode_comp}\footnote{The high compactness case for NFW is included for completeness of the table, since the results for this case presents intricacies as mentioned in \ref{fotnoteNFW}.}. We selected two values in which there is no additional LRs (black line and dashed red line), and a higher compactness regime in which there is an additional pair of LRs in the configuration (solid red line). The black lines show a small deviation from vacuum, with $C=0.1$, and we see that all the three models present similar results to the vacuum case, as already discussed before. 

As we increase the compactness, all the models present a longer decay time with a slower oscillation frequency, as expected due to different asymptotic potential behavior and appearance of additional LRs leading to trapped regions, in the solid red cases. In the regime analyzed, we did not see growing modes in any of these configurations. For the potentials presented in the left panel of Figs.~\ref{FIG:TEHern_Vara0}, \ref{FIG:TENFW}, and \ref{FIG:TEJaffe}, only the Hernquist case appears to clearly exhibit a trapped region in its potential. In addition, the results for high compactness in these plots are presented on a different scale in the right panel of Fig.~\ref{FIG:TEComparison}.

As mentioned above, the presence of stable light rings can indicate the existence of echoes. How clearly these echoes appear, however, depends on additional features, such as the travel time between the two potential barriers, given by
\begin{equation}
\Delta t=\int_{r^-_{\rm LR}}^{r^+_{\rm LR}}\frac{dr}{\sqrt{fh}}.
\end{equation}
To illustrate this point, let us consider the Hernquist model, as presented in Fig.~\ref{FIG:TEHern_Vara0}, at the highest compactness. In this case, the travel time computed above is $\Delta t\approx 1.1\times10^{3}M_{\rm BH}$ or $22a_0$. Therefore, the signal must be evolved for a much longer time in order to observe any echoes. A glimpse of this behavior can be seen in Fig.~\ref{FIG:TEComparison}, where the first reflection is located at approximately $\sim 60 a_0$. Notice that the second peak is larger than the first, since the inner barrier is higher and therefore reflects more efficiently than the outer one.

\begin{figure*}
    \centering
    \includegraphics[width=1\linewidth]{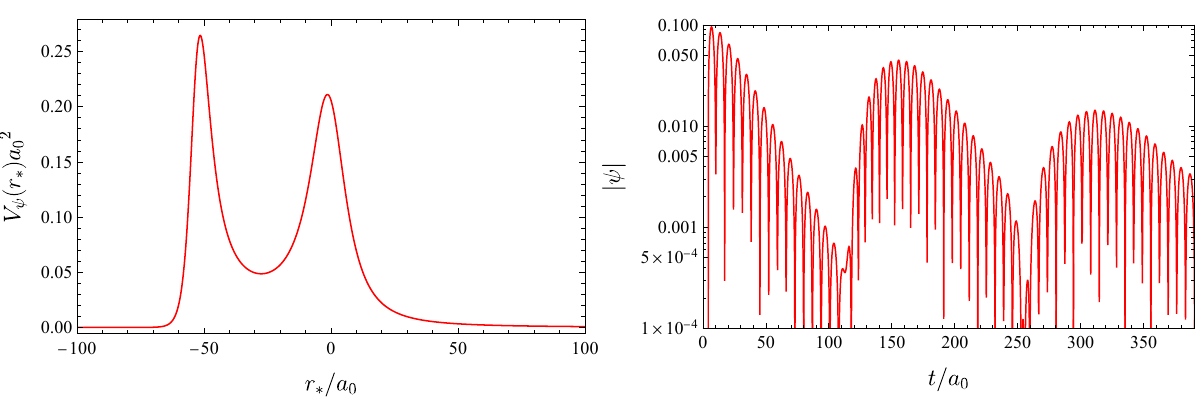}
    \caption{\justifying Evolution of a scalar wavepacket in the Hernquist model with ${C}_{\rm H}=1.6$. The potential (left panel) shows two distinct peaks separated by a trapped region, which allows for clearer echoes. The extracted waveform confirms this behavior (right panel), displaying repeated peaks that slowly decay over time. Notice that the timescale associated with the round trip between the barriers is approximately $100a_0$, which is consistent with the appearance of the second pulse.}
    \label{fig:echoes}
\end{figure*}

To illustrate a cleaner echo phenomenology, let us consider a more extreme case, corresponding to a Hernquist model with ${C}_{\rm H}=1.6$, for which $\Delta t\approx 85 a_0$. The result is visualized in Fig.~\ref{fig:echoes}. The effective potential (shown in the left panel) clearly exhibits two peaks of comparable height, with a trapped region between them. This configuration favors a cleaner echo phenomenology, with consecutive pulses emitted after the first interaction with the potential, as can be seen in the extracted wave (right panel of Fig.~\ref{fig:echoes}). Notice that the scale of the problem is governed by the parameter $a_0$, which is much larger than the black hole scale (similarly to Fig.~\ref{FIG:TEComparison}).

\section{Conclusion} \label{S:VIII}

In this work, we have investigated the impact of spherically symmetric matter distributions on the spacetime geometry and geodesic structure of black holes. Focusing on three widely used astrophysical density profiles---the Hernquist, NFW, and Jaffe models---we constructed self-consistent configurations using an anisotropic-fluid description inspired by Einstein clusters, allowing us to incorporate both the self-gravity of the matter and the presence of a central BH. This framework enabled a systematic analysis of how surrounding matter modifies key relativistic features such as circular geodesics, the ISCO, the light-ring structure, and the associated Lyapunov exponents.

In the low-compactness regime, our analytical expansions revealed that environmental effects generically push the innermost LR outward while drawing the ISCO inward, in agreement with expectations for weakly gravitating halos. These deviations induce small but potentially measurable changes in observables connected to accretion physics, BH imaging, and eikonal QNMs. The Hernquist and NFW profiles exhibit remarkably similar first-order behavior due to their common inner-slope structure, whereas the Jaffe model yields distinct corrections driven by its steeper central density.

In the high-compactness regime, we mapped the full parameter space of each model, identifying regions where new MSCOs, additional pairs of LRs, or secondary horizons emerge. These ultracompact configurations give rise to rich geodesic phenomenology, including patches of unstable timelike circular orbits and the formation of potential wells capable of trapping null geodesics. Our ringdown analysis confirmed that such configurations can imprint characteristic signatures on the waveform, including long-lived quasi-bound states and, in some cases, echo-like modulations associated with multiple potential barriers.

Altogether, our results demonstrate that matter environments—whether diffuse halos or dense dark configurations—can lead to significant departures from the geodesic and vibrational properties of vacuum BHs. As future electromagnetic and gravitational-wave observations reach increasing precision, properly accounting for environmental effects will be essential both for tests of gravity and for disentangling the astrophysical nature of compact objects. The framework developed here provides a step toward this goal and can be extended to rotating backgrounds, alternative equations of state, and dynamical scenarios in forthcoming work.

Regarding the immediate extension of this work concerning the inclusion of rotation, this is an essential issue for realistic modeling of astrophysical BHs and for assessing the detectability of environmental effects with space-based detectors such as LISA \cite{Barausse:2020rsu}. In particular, extreme mass-ratio inspirals (EMRIs) will probe the near-horizon region of massive BHs with unprecedented precision, making them exceptionally sensitive to deviations in the geodesic structure induced by surrounding matter. The presence of additional LRs, modified ISCO locations, or ultracompact configurations could leave imprints on the orbital evolution, phase accumulation, and resonant phenomena characteristic of EMRIs. Since rotation generically couples to both the geometry and the matter distribution, it may enhance or suppress these environmental signatures, potentially generating new families of trapped modes or shifting the onset of strong-field effects. Developing rotating generalizations of the configurations studied here, together with self-consistent EMRI waveform models that incorporate matter-induced corrections, will be crucial for enabling LISA to distinguish genuine signatures of new physics from environmental imprints, and to exploit EMRIs as precision probes of the astrophysical environments surrounding massive BHs.

\section*{Acknowledgments}

C.F.B.M. thanks the Universidad Complutense de Madrid for its kind hospitality during the preparation of this work. C.F.B.M., D.S.F, and M.M.C. acknowledge Coordenação de Aperfeiçoamento de Pessoal de Nível Superior– Brasil (CAPES)– Finance Code 001, Fundação Amazônia de Amparo a Estudos e Pesquisa (FAPESPA), and Conselho Nacional de Desenvolvimento Científico e Tecnológico (CNPq). This work is supported by the Spanish National Grants PID2022-138607NBI00 and CNS2024-154444, funded by MICIU/AEI/10.13039/501100011033 (“PGC Generación de Conocimiento") and FEDER, UE.




\end{document}